\documentstyle[12pt]{article}

\newcommand{\be}{\begin{equation}}
\newcommand{\ee}{\end{equation}}
\newcommand{\bea}{\begin{eqnarray}}
\newcommand{\eea}{\end{eqnarray}}
\newcommand{\ba}{\begin{array}}
\newcommand{\ea}{\end{array}}
\newcommand{\non}{\nonumber}
\newcommand{\PP}[2]{\; P^{(#1)}_{\, #2}}
\newcommand{\pp}[3]{\; p^{(#1,#2)}_{\, #3}}
\newcommand{\eee}[3]{\; e^{(#1,#2)}_{\, #3}}
\newcommand{\bb}[3]{\; b^{(#1,#2)}_{\, #3}}
\newcommand{\bbi}[3]{\; {b^{-\,}}^{(#1,#2)}_{\, #3}}
\newcommand{\bbp}[3]{\; {b^{+\,}}^{(#1,#2)}_{\, #3}}
\newcommand{\bbpm}[3]{\; {b^{\pm\,}}^{(#1,#2)}_{\, #3}}
\newcommand{\fff}[2]{f^{(#1)}(#2)}
\newcommand{\ggg}[2]{g^{(#1)}(#2)}
\newcommand{\EE}[1]{E_{\, #1}}
\newcommand{\BB}[1]{B_{\, #1}}
\newcommand{\BBi}[1]{B^{-1}_{\, #1}}
\newcommand{\BBpm}[1]{B^{\pm}_{\, #1}}
\newcommand{\id}{\mbox{$\; I\; $}}
\newcommand{\om}[1]{\omega^{(#1)}}
\newcommand{\sq}[1]{\sqrt{Q^{(#1)}}}
\newcommand{\slu}{\frac{\sin(\lambda - u)}{\sin(\lambda)}}
\newcommand{\su}{\frac{\sin(u)}{\sin(\lambda)}}
\newcommand{\sslu}[2]{\frac{\sin(#1\ol{\lambda} - u)}{\sin(#2\ol{\lambda})}}
\newcommand{\ssu}[1]{\frac{\sin(u)}{\sin(#1\ol{\lambda})}}
\newcommand{\slulu}[4]{\frac{\sin(#1\ol{\lambda} - u)\:
                             \sin(#2\ol{\lambda} - u)}
                       {\sin(#3\ol{\lambda})\:\sin(#4\ol{\lambda})}}
\newcommand{\sulu}[3]{\frac{\sin(u)\:\sin(#1\ol{\lambda} - u)}
                      {\sin(#2\ol{\lambda})\:\sin(#3\ol{\lambda})}}
\newcommand{\zr}[1]{\mbox{\hspace*{#1em}}}
\newcommand{\CC}{\mbox{\zr{0.1}\rule{0.04em}{1.6ex}\zr{-0.30}{\sf C}}}
\newcommand{\ZZ}{\mbox{\sf Z\zr{-0.45}Z}}
\newcommand{\VM}[3]{$\mbox{#1}^{(#3)}_{#2}$}
\newcommand{\ul}{\underline}
\newcommand{\ol}{\overline}
\newcommand{\mbfa}{\mbox{\protect\boldmath $a$}}
\newcommand{\mbfap}{\mbox{\protect\boldmath $a'$}}
\newcommand{\mbfb}{\mbox{\protect\boldmath $b$}}
\newcommand{\mbfs}{\mbox{\protect\boldmath $s$}}
\newcommand{\mbfsp}{\mbox{\protect\boldmath $s'$}}
\newcommand{\eqstop}[1]{\hspace*{4mm} #1}
\newcommand{\rrr}[2]{\rule[-#1\unitlength]{0in}{#2\unitlength}}
\newcommand{\ww}[5]{\renewcommand{\arraystretch}{0.8}
\mbox{$W\left(\left.\!\ba{c@{\;\;}c}
\mbox{\footnotesize $#4$} & \mbox{\footnotesize $#3$} \\
\mbox{\footnotesize $#1$} & \mbox{\footnotesize $#2$} \ea \right|\, #5 \,
\right)$} \renewcommand{\arraystretch}{1}}
\newcommand{\WW}[5]{\renewcommand{\arraystretch}{0.6}
\mbox{$W\left(\left.\!\ba{c@{\;}c@{\;}c}
& \mbox{\footnotesize $#3$} & \\
\mbox{\footnotesize $#4$} & & \mbox{\footnotesize $#2$} \\
& \mbox{\footnotesize $#1$} & \ea \right|\, #5 \,
\right)$} \renewcommand{\arraystretch}{1}}
\newtheorem{theorem}{\underline{Theorem:}}

\newtheorem{conjecture}{\underline{Conjecture:}}

\thicklines
\def\punit#1{\hspace{#1\unitlength}}

\def\braidcc#1#2#3#4#5#6#7#8{\rule[-4\unitlength]{0in}{8\unitlength}
\begin{picture}(0,0)(-#1,-#2)
\put(0,0){\line(1,1){4}}
\put(0,4){\line(1,-1){1.5}}
\put(4,0){\line(-1,1){1.5}}
\put(0,-1){\makebox(0,0)[t]{\footnotesize \mbox{$#3$}\rule[0pt]{0pt}{7pt}}}
\put(4,-1){\makebox(0,0)[t]{\footnotesize \mbox{$#4$}\rule[0pt]{0pt}{7pt}}}
\put(-1,0){\makebox(0,0)[l]{\mbox{$\scriptstyle #5$}}}
\put(5,0){\makebox(0,0)[r]{\mbox{$\scriptstyle #6$}}}
\put(-1,4){\makebox(0,0)[l]{\mbox{$\scriptstyle #7$}}}
\put(5,4){\makebox(0,0)[r]{\mbox{$\scriptstyle #8$}}}
\end{picture}}

\def\invbraidcc#1#2#3#4#5#6#7#8{\rule[-4\unitlength]{0in}{8\unitlength}
\begin{picture}(0,0)(-#1,-#2)
\put(0,4){\line(1,-1){4}}
\put(0,0){\line(1,1){1.5}}
\put(4,4){\line(-1,-1){1.5}}
\put(0,-1){\makebox(0,0)[t]{\footnotesize \mbox{$#3$}\rule[0pt]{0pt}{7pt}}}
\put(4,-1){\makebox(0,0)[t]{\footnotesize \mbox{$#4$}\rule[0pt]{0pt}{7pt}}}
\put(-1,0){\makebox(0,0)[l]{\mbox{$\scriptstyle #5$}}}
\put(5,0){\makebox(0,0)[r]{\mbox{$\scriptstyle #6$}}}
\put(-1,4){\makebox(0,0)[l]{\mbox{$\scriptstyle #7$}}}
\put(5,4){\makebox(0,0)[r]{\mbox{$\scriptstyle #8$}}}
\end{picture}}

\def\monoidcc#1#2#3#4#5#6#7#8{\rule[-4\unitlength]{0in}{8\unitlength}
\begin{picture}(0,0)(-#1,-#2)
\put(2,0){\oval(4,3.5)[t]}
\put(2,4){\oval(4,3.5)[b]}
\put(0,-1){\makebox(0,0)[t]{\footnotesize \mbox{$#3$}\rule[0pt]{0pt}{7pt}}}
\put(4,-1){\makebox(0,0)[t]{\footnotesize \mbox{$#4$}\rule[0pt]{0pt}{7pt}}}
\put(-1,0){\makebox(0,0)[l]{\mbox{$\scriptstyle #5$}}}
\put(5,0){\makebox(0,0)[r]{\mbox{$\scriptstyle #6$}}}
\put(-1,4){\makebox(0,0)[l]{\mbox{$\scriptstyle #7$}}}
\put(5,4){\makebox(0,0)[r]{\mbox{$\scriptstyle #8$}}}
\end{picture}}

\def\project#1#2#3#4{\rule[-4\unitlength]{0in}{8\unitlength}
\begin{picture}(0,0)(-#1,-#2)
\put(0,0){\line(0,1){4}}
\linethickness{2pt}
\put(0,0){\line(0,1){2}}
\thicklines
\put(0,-1){\makebox(0,0)[t]{\footnotesize \mbox{$#3$}\rule[0pt]{0pt}{7pt}}}
\put(1,4){\makebox(0,0)[r]{\mbox{$\scriptstyle #4$}}}
\end{picture}}

\def\vertcr#1#2#3#4#5{\rule[-4\unitlength]{0in}{8\unitlength}
\begin{picture}(0,0)(-#1,-#2)
\put(0,0){\line(0,1){4}}
\put(0,-1){\makebox(0,0)[t]{\footnotesize \mbox{$#3$}\rule[0pt]{0pt}{7pt}}}
\put(1,0){\makebox(0,0)[r]{\mbox{$\scriptstyle #4$}}}
\put(1,4){\makebox(0,0)[r]{\mbox{$\scriptstyle #5$}}}
\end{picture}}

\def\vertcl#1#2#3#4#5{\rule[-4\unitlength]{0in}{8\unitlength}
\begin{picture}(0,0)(-#1,-#2)
\put(0,0){\line(0,1){4}}
\put(0,-1){\makebox(0,0)[t]{\footnotesize \mbox{$#3$}\rule[0pt]{0pt}{7pt}}}
\put(-1,0){\makebox(0,0)[l]{\mbox{$\scriptstyle #4$}}}
\put(-1,4){\makebox(0,0)[l]{\mbox{$\scriptstyle #5$}}}
\end{picture}}

\def\vertf#1#2#3{\rule[-4\unitlength]{0in}{8\unitlength}
\linethickness{2pt}
\begin{picture}(0,0)(-#1,-#2)
\put(0,0){\line(0,1){4}}
\put(0,-1){\makebox(0,0)[t]{\footnotesize \mbox{$#3$}\rule[0pt]{0pt}{7pt}}}
\end{picture}
\thicklines}

\def\dface#1#2#3#4#5#6#7{
\setlength{\unitlength}{.0943in}
\rule[-4\unitlength]{0in}{8\unitlength}
\begin{picture}(6,4)(-#6,-#7)
\put(1.5,0.5){\line(1,-1){3}}
\put(4.5,3.5){\line(1,-1){3}}
\put(4.5,3.5){\line(-1,-1){3}}
\put(7.5,0.5){\line(-1,-1){3}}
\put(4.5,-3){\makebox(0,0)[t]{\footnotesize \mbox{$#1$}\rule[-2pt]{0pt}{9pt}}}
\put(8,0.5){\makebox(0,0)[l]{\footnotesize \mbox{$#2$}\rule[-2pt]{0pt}{9pt}}}
\put(4.5,4){\makebox(0,0)[b]{\footnotesize \mbox{$#3$}\rule[-2pt]{0pt}{9pt}}}
\put(1,0.5){\makebox(0,0)[r]{\footnotesize \mbox{$#4$}\rule[-2pt]{0pt}{9pt}}}
\put(4.5,0.5){\makebox(0,0){\footnotesize \mbox{$#5$}}}
\end{picture}
\setlength{\unitlength}{.1in}}

\def\dvertex#1#2#3#4#5#6#7{\rule[-3.5\unitlength]{0in}{7\unitlength}
\begin{picture}(4,4)(-#6,-#7)
\put(0,-2){\line(1,1){4}}
\put(4,-2){\line(-1,1){4}}
\put(0,-2.5){\makebox(0,0)[t]{\footnotesize \mbox{$#1$}\rule[-2pt]{0pt}{9pt}}}
\put(4,-2.5){\makebox(0,0)[t]{\footnotesize \mbox{$#2$}\rule[-2pt]{0pt}{9pt}}}
\put(4,2.5){\makebox(0,0)[b]{\footnotesize \mbox{$#3$}\rule[-2pt]{0pt}{9pt}}}
\put(0,2.5){\makebox(0,0)[b]{\footnotesize \mbox{$#4$}\rule[-2pt]{0pt}{9pt}}}
\put(2,-0.5){\makebox(0,0)[t]{\mbox{$\scriptstyle #5$}}}
\put(2,-0.9){\makebox(0,0)[t]{\footnotesize \mbox{$\smile$}}}
\end{picture}}

\begin{document}


\begin{center}
{\LARGE\bf Multi-Colour Braid-Monoid Algebras} \\[8mm]
{\large\sc Uwe Grimm \hspace*{1pt} and \hspace*{1pt} Paul A. Pearce} \\[2mm]
{\footnotesize Department of Mathematics, University of Melbourne,\\
        Parkville, Victoria 3052, Australia} \\[6mm]
March 1993 \\[8mm]
\end{center}


\begin{quote}
{\small\sf
 We define multi-colour generalizations of braid-monoid algebras
 and present explicit matrix representations which are related to
 two-dimensional exactly solvable lattice models of statistical mechanics.
 In particular, we show that the two-colour
 braid-monoid algebra describes the
 Yang-Baxter algebra of the  critical dilute A--D--E models which
 were recently introduced by Warnaar, Nienhuis, and Seaton
 as well as by Roche.
 These and other solvable models related to
 dense and dilute loop models are discussed in detail and it
 is shown that the solvability is a direct consequence of the algebraic
 structure.
 It is conjectured that the Yang-Baxterization of general multi-colour
 braid-monoid algebras will lead to the construction of
 further solvable lattice models.}
\end{quote}
\vspace*{2mm}



\section{Introduction}
\setcounter{equation}{0}

The study of Yang-Baxter equations \cite{Jimbo89}
has revealed a rich underlying algebraic
structure in integrable systems with applications
ranging from statistical mechanics to
knot theory. Most well known among these algebraic
structures is the quantum group.
However, Yang-Baxter algebras are intimately connected
to a number of other algebraic
structures, most notably, the braid group \cite{YangGe89}
and the Temperley-Lieb
\cite{TempLieb} and Hecke algebras \cite{Martin91}.
The braid and Temperley-Lieb or monoid
\cite{Kauffman} operators were first combined into a single
algebra in 1987 by Birman and Wenzl \cite{BirWen} and independently
by Murakami \cite{Mura}. Subsequently, generalized braid-monoid algebras
were introduced by Wadati, Deguchi and Akutsu \cite{WadDegAku}.
On the one hand, these algebras are
related to certain two-dimensional exactly solvable
lattice models. On the other hand, the
braid-monoid algebras admit a simple diagrammatic
interpretation which points to the connections with
the isotopy of knots and links. In this paper, we consider a
generalization of the braid-monoid
algebras where each strand or string of a link is
assigned a colour. Some representations of
these extended algebras correspond to the
Yang-Baxter algebras of new critical solvable
lattice models recently obtained by the
Amsterdam group \cite{WarNie}. In particular, the
two colour algebra is related to the dilute
A--D--E models \cite{WarNieSea,Roche}.

The paper is organized as follows.
We begin by defining the multi-colour braid-monoid algebra in sec.~2.
A graphical interpretation is presented and a generalized notion of
crossing symmetry is introduced.
In sec.~3, we consider matrix representations of the multi-colour algebra
and give explicit expressions for different types of representations.
Sec.~4 concentrates on the two-colour case and the
relation to exactly solvable models. The Yang-Baxter algebra
of several RSOS models, vertex models, and mixed vertex-RSOS models
is shown to be described by the two-colour algebra.
Our main point, however, is that the algebraic structure is actually
sufficient to guarantee that the Yang-Baxter equations are
satisfied and hence yields the solvability of the models.
We conclude by summarizing our results and give an outlook
on possible consequences and further investigations.
In particular, we conjecture that
the general multi-colour braid-monoid can be
Yang-Baxterized \cite{Jones} to obtain new
exactly solvable lattice models. Before proceeding, we
point out that coloured braids and links have been
considered previously by Akutsu, Deguchi
and Wadati \cite{AkuDeg91,DegAku91a,Deg,DegAku91b,AkuDegWad92}.
However, the developments of these papers are
unrelated to the present paper.



\section{Definition of the Multi-Colour Algebra}
\setcounter{equation}{0}

The $(n+1)$-string $m$-colour braid-monoid algebra is the
associative algebra generated by the unit element $\id$
(i.e., $\id x = x\id = x$ for all elements $x$ of
the algebra),
central elements $\sq{\alpha}$ and $\om{\alpha}$
($1\leq\alpha\leq m$)
together with three sets of generators
\begin{itemize}
\item $m (n+1)$ ``projectors'' $\PP{\alpha}{j}$
      ($1\leq j\leq n+1$, $1\leq\alpha\leq m$)
\item $m^2 n$ ``coloured Temperley-Lieb operators''
      $\eee{\alpha}{\beta}{j}$
\item $m^2 n$ ``coloured braids''
      $\bbp{\alpha}{\beta}{j}=\bb{\alpha}{\beta}{j}$
      and ``coloured inverse braids'' $\bbi{\alpha}{\beta}{j}$
\end{itemize}
which fulfill the following list of relations
\begin{enumerate}
\item \ul{projector relations:}
\bea
\PP{\beta}{j}\PP{\alpha}{j} & = & \delta_{\alpha,\beta}\PP{\alpha}{j}\non\\
\PP{\beta}{k}\PP{\alpha}{j} & = & \PP{\alpha}{j} \PP{\beta}{k}
\hspace*{1cm}\mbox{for $j\neq k$} \label{projrel}\\
\sum_{\alpha=1}^{m}\PP{\alpha}{j} & = & \id\non
\eea
\item \ul{braid relations:}
\bea
\bbi{\beta}{\alpha}{j} \bb{\alpha}{\beta}{j} & = &
\bb{\beta}{\alpha}{j} \bbi{\alpha}{\beta}{j} \;\; = \;\;
\PP{\alpha}{j} \PP{\beta}{j+1} \non \\
\bb{\gamma}{\delta}{k} \bb{\alpha}{\beta}{j} & = &
\bb{\alpha}{\beta}{j} \bb{\gamma}{\delta}{k}
\hspace{1cm} \mbox{for $|j-k|>1$} \label{braidrel} \\
\bb{\gamma}{\alpha}{j+1} \bb{\gamma}{\beta}{j} \bb{\alpha}{\beta}{j+1} & = &
\bb{\alpha}{\beta}{j} \bb{\gamma}{\beta}{j+1} \bb{\gamma}{\alpha}{j} \non
\eea
\item \ul{Temperley-Lieb (monoid) relations:}
\bea
\eee{\beta}{\gamma}{j} \eee{\alpha}{\beta}{j} & = &
\sq{\beta} \eee{\alpha}{\gamma}{j} \non \\
\eee{\gamma}{\delta}{k} \eee{\alpha}{\beta}{j} & = &
\eee{\alpha}{\beta}{j} \eee{\gamma}{\delta}{k}
\hspace{1cm} \mbox{for $|j-k|>1$} \label{TLA} \\
\eee{\beta}{\gamma}{j} \eee{\beta}{\beta}{j\pm1}
\eee{\alpha}{\beta}{j} & = &
\eee{\alpha}{\gamma}{j} \PP{\beta}{j+(1\pm 3)/2} \non
\eea
\item \ul{``twist'' relations:}
\bea
\eee{\gamma}{\delta}{k} \bb{\alpha}{\beta}{j} & = &
\bb{\alpha}{\beta}{j} \eee{\gamma}{\delta}{k}
\hspace{1cm} \mbox{for $|j-k|>1$} \non \\
\bb{\beta}{\beta}{j} \eee{\alpha}{\beta}{j} & = &
\om{\beta} \eee{\alpha}{\beta}{j} \label{twist} \\
\eee{\alpha}{\beta}{j} \bb{\alpha}{\alpha}{j} & = &
\om{\alpha} \eee{\alpha}{\beta}{j} \non
\eea
\item \ul{braid-monoid relations:}
\bea
\bb{\gamma}{\beta}{j+1} \bb{\gamma}{\beta}{j} \eee{\alpha}{\beta}{j+1}
& = &
\eee{\alpha}{\beta}{j} \bb{\gamma}{\alpha}{j+1} \bb{\gamma}{\alpha}{j}
\;\; = \;\;
\eee{\gamma}{\beta}{j} \eee{\alpha}{\gamma}{j+1} \non \\
\bb{\beta}{\gamma}{j-1} \bb{\beta}{\gamma}{j} \eee{\alpha}{\beta}{j-1}
& = &
\eee{\alpha}{\beta}{j} \bb{\alpha}{\gamma}{j-1} \bb{\alpha}{\gamma}{j}
\;\; = \;\;
\eee{\gamma}{\beta}{j} \eee{\alpha}{\gamma}{j-1} \label{BBE}
\eea
\item \ul{compatibility relations between projectors and braids:}
\bea
\bbpm{\beta}{\gamma}{k}\PP{\alpha}{j} & = &
\PP{\alpha}{j} \bbpm{\beta}{\gamma}{k}
\hspace*{1cm} \mbox{for $j\neq k,k+1$} \non \\
\bbpm{\beta}{\gamma}{j} \PP{\alpha}{j} & = &
\PP{\alpha}{j+1} \bbpm{\beta}{\gamma}{j} \;\; = \;\;
\delta_{\alpha,\beta} \bbpm{\beta}{\gamma}{j} \label{compb} \\
\bbpm{\beta}{\gamma}{j} \PP{\alpha}{j+1} & = &
\PP{\alpha}{j} \bbpm{\beta}{\gamma}{j} \;\; = \;\;
\delta_{\alpha,\gamma} \bbpm{\beta}{\gamma}{j} \non
\eea
\item \ul{compatibility relations between projectors and monoids:}
\bea
\eee{\beta}{\gamma}{k} \PP{\alpha}{j} & = &
\PP{\alpha}{j} \eee{\beta}{\gamma}{k}
\hspace*{1cm} \mbox{for $j\neq k,k+1$} \non \\
\eee{\beta}{\gamma}{j} \PP{\alpha}{j} & = &
\eee{\beta}{\gamma}{j} \PP{\alpha}{j+1} \;\; = \;\;
\delta_{\alpha,\beta} \eee{\beta}{\gamma}{j} \label{compe} \\
\PP{\alpha}{j} \eee{\beta}{\gamma}{j} & = &
\PP{\alpha}{j+1} \eee{\beta}{\gamma}{j} \;\; = \;\;
\delta_{\alpha,\gamma} \eee{\beta}{\gamma}{j} \non
\eea
\end{enumerate}


\subsection{Graphical Interpretation}

The algebra defined above allows a graphical presentation acting
on $n\! +\! 1$ strings which themselves can be thought of
as composite objects consisting of $m$ ``coloured'' strings each.
In the pictures below, we are going to represent the ``full''
(``uncoloured'') string by a bold (thick) line whereas strings with a
colour are represented by a thin line carrying the colour index.


Our pictorial view of the generators looks as follows.
The projectors $\PP{\alpha}{j}$ are represented by
\be
\PP{\alpha}{j}\punit3  = \punit4
  \vertf{0}{-1.5}{1}\vertf{4}{-1.5}{2}\punit6\ldots\punit2
  \vertf{0}{-1.5}{j\! -\! 1}\project{4}{-1.5}{j}{\alpha}
  \vertf{8}{-1.5}{j\! +\! 1}
  \vertf{12}{-1.5}{j\! +\! 2}\punit{14}\ldots\punit2
  \vertf{0}{-1.5}{n}\vertf{4}{-1.5}{n\! +\! 1}\punit8
\ee
or simpler by
\be
\PP{\alpha}{j}\punit3 = \punit4
  \vertf{0}{-1.5}{1}\vertf{4}{-1.5}{2}\punit6\ldots\punit2
  \vertf{0}{-1.5}{j\! -\! 1}\vertcl{4}{-1.5}{j}{\alpha}{\alpha}
  \vertf{8}{-1.5}{j\! +\! 1}
  \vertf{12}{-1.5}{j\! +\! 2}\punit{14}\ldots\punit2
  \vertf{0}{-1.5}{n}\vertf{4}{-1.5}{n\! +\! 1}\punit8
\ee
since we can always multiply with the identity
\be
\id\punit3 = \punit4
  \vertf{0}{-1.5}{1}\vertf{4}{-1.5}{2}\punit6\ldots\punit2
  \vertf{0}{-1.5}{j\! -\! 1}
  \vertf{4}{-1.5}{j}
  \vertf{8}{-1.5}{j\! +\! 1}\punit{10}\ldots\punit2
  \vertf{0}{-1.5}{n}\vertf{4}{-1.5}{n\! +\! 1}\punit8
\ee
The coloured braids, inverse braids, and monoids correspond
to the diagrams
\bea
\bb{\alpha}{\beta}{j}\punit2 & = &
 \punit3\vertf{0}{-1.5}{1}\vertf{4}{-1.5}{2}\punit6\ldots\punit2
  \vertf{0}{-1.5}{j\! -\! 1}
  \braidcc{4}{-1.5}{j}{j\! +\! 1}{\alpha}{\beta}{\beta}{\alpha}
  \vertf{12}{-1.5}{j\! +\! 2}\punit{14}\ldots\punit2
  \vertf{0}{-1.5}{n}\vertf{4}{-1.5}{n\! +\! 1}\punit8 \\
\bbi{\alpha}{\beta}{j}\punit2 & = &
 \punit3\vertf{0}{-1.5}{1}\vertf{4}{-1.5}{2}\punit6\ldots\punit2
  \vertf{0}{-1.5}{j\! -\! 1}
  \invbraidcc{4}{-1.5}{j}{j\! +\! 1}{\alpha}{\beta}{\beta}{\alpha}
  \vertf{12}{-1.5}{j\! +\! 2}\punit{14}\ldots\punit2
  \vertf{0}{-1.5}{n}\vertf{4}{-1.5}{n\! +\! 1}\punit8 \\
\eee{\alpha}{\beta}{j}\punit2 & = &
 \punit3\vertf{0}{-1.5}{1}\vertf{4}{-1.5}{2}\punit6\ldots\punit2
  \vertf{0}{-1.5}{j\! -\! 1}
  \monoidcc{4}{-1.5}{j}{j\! +\! 1}{\alpha}{\alpha}{\beta}{\beta}
  \vertf{12}{-1.5}{j\! +\! 2}\punit{14}\ldots\punit2
  \vertf{0}{-1.5}{n}\vertf{4}{-1.5}{n\! +\! 1}\punit8
\eea


Multiplication corresponds to concatenation of pictures
where we use the convention that the product $A\cdot B$ corresponds
to glueing the picture for $A$ above the one for $B$
(hence if one thinks of the generators as operators acting on
states (to the right) then the ``time-direction'' points upwards).
Any picture with an incompatible matching of colours is zero.
Diagrams which can be transformed into each other by
continuous deformations of strings
(without affecting the colours of strings, of course)
are equivalent. All the defining
relations of the multi-colour braid-monoid algebra can
be visualized in this way.

To illustrate this, we present pictorial versions of
the main defining relations as examples.
The projector relations (\ref{projrel})
and the compatibility relations
(\ref{compb}) and (\ref{compe})
follow from the requirement of matching colours alone.
Also, the commutativity of operators
acting on different strings is represented in
the diagrams in an obvious way.
For the remaining relations, the corresponding pictures
are given in the sequel.

The braid relations (\ref{braidrel}) give rise to the
following diagrams
\be
\rrr{6}{12}
\braidcc{0}{-4}{j}{j\! +\! 1}{\alpha}{\beta}{\beta}{\alpha}
\invbraidcc{0}{0}{}{}{}{}{\alpha}{\beta}
\punit9  = \punit5
\invbraidcc{0}{-4}{j}{j\! +\! 1}{\alpha}{\beta}{\beta}{\alpha}
\braidcc{0}{0}{}{}{}{}{\alpha}{\beta}
\punit9 = \punit5
\vertcl{0}{-4}{j}{\alpha}{}\vertcl{0}{0}{}{}{\alpha}\punit4
\vertcr{0}{-4}{j\! +\! 1}{\beta}{}\vertcr{0}{0}{}{}{\beta}\punit2
\ee
\be
\rrr{9}{18}
\vertcl{0}{-6}{j}{\gamma}{}
\vertcl{0}{2}{}{}{\beta}
\braidcc{0}{-2}{}{}{}{}{}{}\punit4
\braidcc{0}{-6}{j\! +\! 1}{j\! +\! 2}{\alpha}{\beta}{}{}
\braidcc{0}{2}{}{}{}{}{\alpha}{\gamma}\punit4\vertcr{0}{-2}{}{}{}
\punit5 = \punit5
\braidcc{0}{-6}{j}{j\! +\! 1}{\gamma}{\alpha}{}{}
\vertcr{0}{-2}{}{}{}
\braidcc{0}{2}{}{}{}{}{\beta}{\alpha}\punit4
\braidcc{0}{-2}{}{}{}{}{}{}\punit4
\vertcr{0}{-6}{j\! +\! 2}{\beta}{}\vertcr{0}{2}{}{}{\gamma}
\ee
where here and in what follows we only show the relevant part.

The multi-colour Temperley-Lieb relations (\ref{TLA}) are
\be
\rrr{5}{12}
\monoidcc{0}{-3}{j}{j\! +\! 1}{\alpha}{\alpha}{\beta}{\beta}
\monoidcc{0}{1}{}{}{}{}{\gamma}{\gamma}
\punit9 = \punit5
\sq{\beta}\punit4
\monoidcc{0}{-1}{j}{j\! +\! 1}{\alpha}{\alpha}{\gamma}{\gamma}
\ee
\be
\rrr{8}{16}
\monoidcc{0}{-6}{j}{j\! +\! 1}{\alpha}{\alpha}{}{}
\vertcr{0}{-2}{}{}{}
\monoidcc{0}{2}{}{}{}{}{\gamma}{\gamma}\punit4
\monoidcc{0}{-2}{}{}{}{}{}{}\punit4
\vertcr{0}{-6}{j\! +\! 2}{\beta}{}\vertcr{0}{2}{}{}{\beta}
\punit5 = \punit5
\monoidcc{0}{-2}{j}{j\! +\! 1}{\alpha}{\alpha}{\gamma}{\gamma}
\punit8\vertcr{0}{-2}{j\! +\! 2}{\beta}{\beta}
\ee
\be
\rrr{9}{18}
\monoidcc{0}{-2}{}{}{}{}{}{}
\vertcl{0}{-6}{j\! -\! 1}{\beta}{}\vertcl{0}{2}{}{}{\beta}
\punit4
\monoidcc{0}{-6}{j}{j\! +\! 1}{\alpha}{\alpha}{}{}
\monoidcc{0}{2}{}{}{}{}{\gamma}{\gamma}\punit4
\vertcr{0}{-2}{}{}{}
\punit5 = \punit5
\vertcl{0}{-2}{j\! -\! 1}{\beta}{\beta}\non
\punit4
\monoidcc{0}{-2}{j}{j\! +\! 1}{\alpha}{\alpha}{\gamma}{\gamma}
\punit4
\ee
which means that a closed loop of colour $\alpha$
in a diagram can be replaced by the diagram
without the loop multiplied by a factor $\sq{\alpha}$.

The ``twist'' relations (\ref{twist}) have the following
pictorial interpretation
\bea
\rrr{5}{12}
\monoidcc{0}{-3}{j}{j\! +\! 1}{\alpha}{\alpha}{\beta}{\beta}
\braidcc{0}{1}{}{}{}{}{\beta}{\beta}
\punit8 & = & \punit4\om{\beta}\punit4
\monoidcc{0}{-1}{j}{j\! +\! 1}{\alpha}{\alpha}{\beta}{\beta} \\
\rrr{5}{12}
\braidcc{0}{-3}{j}{j\! +\! 1}{\alpha}{\alpha}{}{}
\monoidcc{0}{1}{}{}{\alpha}{\alpha}{\beta}{\beta}
\punit8 & = & \punit4\om{\alpha}\punit4
\monoidcc{0}{-1}{j}{j\! +\! 1}{\alpha}{\alpha}{\beta}{\beta}
\eea
i.e., undoing a ``twist'' in a string of colour $\alpha$
produces a factor $\om{\alpha}$.

Finally, the braid-monoid relations (\ref{BBE}) look as
follows
\bea
\rrr{9}{18}
\braidcc{0}{-2}{}{}{}{}{}{}
\vertcl{0}{-6}{j}{\gamma}{}\vertcl{0}{2}{}{}{\beta}
\punit4
\monoidcc{0}{-6}{j\! +\! 1}{j\! +\! 2}{\alpha}{\alpha}{}{}
\braidcc{0}{2}{}{}{}{}{\beta}{\gamma}\punit4
\vertcr{0}{-2}{}{}{}
\punit4 & = & \punit4
\braidcc{0}{-6}{j}{j\! +\! 1}{\gamma}{\alpha}{}{}
\vertcr{0}{-2}{}{}{}{}
\monoidcc{0}{2}{}{}{}{}{\beta}{\beta}
\punit4
\braidcc{0}{-2}{}{}{}{}{}{}\punit4
\vertcr{0}{2}{}{}{\gamma}
\vertcr{0}{-6}{j\! +\! 2}{\alpha}{}
\punit5 = \punit5
\vertcl{0}{-4}{j}{\gamma}{}
\monoidcc{0}{0}{}{}{}{}{\beta}{\beta}
\punit4
\monoidcc{0}{-4}{j\! +\! 1}{j\! +\! 2}{\alpha}{\alpha}{}{}
\punit4\vertcr{0}{0}{}{}{\gamma} \\
\rrr{9}{18}
\monoidcc{0}{-6}{j\! -\! 1}{j}{\alpha}{\alpha}{}{}
\vertcr{0}{-2}{}{}{}
\braidcc{0}{2}{}{}{}{}{\gamma}{\beta}
\punit4
\braidcc{0}{-2}{}{}{}{}{}{}
\punit4
\vertcr{0}{-6}{j\! +\! 1}{\gamma}{}
\vertcr{0}{2}{}{}{\beta}
\punit4 & = & \punit4
\braidcc{0}{-2}{}{}{}{}{}{}
\vertcl{0}{-6}{j\! - \!1}{\alpha}{}
\vertcl{0}{2}{}{}{\gamma}
\punit4
\braidcc{0}{-6}{j}{j\! +\! 1}{\alpha}{\gamma}{}{}
\monoidcc{0}{2}{}{}{}{}{\beta}{\beta}
\punit4
\vertcr{0}{-2}{}{}{}
\punit5 = \punit5
\monoidcc{0}{-4}{j\! -\! 1}{j}{\alpha}{\alpha}{}{}
\vertcl{0}{0}{}{}{\gamma}
\punit4
\monoidcc{0}{0}{}{}{}{}{\beta}{\beta}
\punit4\vertcr{0}{-4}{j\! +\! 1}{\gamma}{}
\eea

This completes the list of defining relations. The
pictorial presentation is useful since it allows to
simplify products of operators in a fast and easy way.
More importantly, however, it shows that
there is a close connection to the theory of
coloured knots and links and it should be
possible to derive invariants for the coloured objects
in the same way one obtains invariants for
knots and links from the usual braid-monoid algebra
(see e.g.~\cite{WadDegAku} and \cite{CheGeLiuXue}).


\subsection{Generalized Crossing Symmetry}

Before we commence to investigate representations of the multi-colour
algebra, let us say a few words on a generalization of the
so-called crossing symmetry
which is actually a built-in feature of the algebra.
The notion of crossing symmetry originates in scattering theory
and was introduced in the context of statistical mechanics models
via the close relation of exactly solvable two-dimensional
statistical models and completely integrable quantum systems
in one dimension (where the Yang-Baxter equations
guarantee the factorization of S-matrices,
see e.g.~\cite{WadDegAku}),
compare the discussion of crossing
symmetry in solvable models in sec.~4.1 below.
We do not attempt to formulate
crossing in full generality but we rather present
some simple examples which clarify our notion of crossing.
Since the colours of strings are not altered by crossing
we neglect all the colour indices of the strings for the moment.
However, one should bear in mind that each string is meant to
have a definite colour in the pictures below (or that
one anyhow looks at the one-colour case).
This restriction is necessary here since we do not have an
interpretation for crossing of the projectors.

To explain what we mean by generalized crossing,
we use the graphical interpretation of
the algebra. Essentially, if one has any (sub-) diagram
(symbolized by a rectangular box in the figure below)
with $\ell$ in- and outgoing (coloured) strings,
\be
\begin{picture}(28,18)(-2,-7)
\put(0,-4){
\begin{picture}(0,0)(0,0)
\put(0,0){\line(0,1){4}}
\put(4,0){\line(0,1){4}}
\put(8,0){\line(0,1){4}}
\put(8,0){\makebox(12,0)[c]{\ldots}}
\put(20,0){\line(0,1){4}}
\put(24,0){\line(0,1){4}}
\put(0,-1){\makebox(0,0)[t]{\footnotesize \mbox{$1$}\rule[0pt]{0pt}{7pt}}}
\put(4,-1){\makebox(0,0)[t]{\footnotesize \mbox{$2$}\rule[0pt]{0pt}{7pt}}}
\put(8,-1){\makebox(0,0)[t]{\footnotesize \mbox{$3$}\rule[0pt]{0pt}{7pt}}}
\put(20,-1){\makebox(0,0)[t]{\footnotesize
                             \mbox{$\ell\! -\! 1$}\rule[0pt]{0pt}{7pt}}}
\put(24,-1){\makebox(0,0)[t]{\footnotesize \mbox{$\ell$}\rule[0pt]{0pt}{7pt}}}
\end{picture}}
\put(-2,0){\line(1,0){28}}
\put(-2,0){\line(0,1){4}}
\put(-2,4){\line(1,0){28}}
\put(26,0){\line(0,1){4}}
\put(0,4){
\begin{picture}(0,0)(0,0)
\put(0,0){\line(0,1){4}}
\put(4,0){\line(0,1){4}}
\put(8,0){\line(0,1){4}}
\put(8,4){\makebox(12,0)[c]{\ldots}}
\put(20,0){\line(0,1){4}}
\put(24,0){\line(0,1){4}}
\put(0,5){\makebox(0,0)[b]{\footnotesize \mbox{$1'$}}}
\put(4,5){\makebox(0,0)[b]{\footnotesize \mbox{$2'$}}}
\put(8,5){\makebox(0,0)[b]{\footnotesize \mbox{$3'$}}}
\put(20,5){\makebox(0,0)[b]{\footnotesize \mbox{$(\ell\! -\! 1)'$}}}
\put(24,5){\makebox(0,0)[b]{\footnotesize \mbox{$\ell'$}}}
\end{picture}}
\end{picture}
\ee
applying one generalized crossing step (``crossing generator'')
means that one changes the interpretation
of two strings as follows:
\be
\begin{picture}(50,18)(0,0)
\put(0,0){
\begin{picture}(20,18)(-2,-7)
\put(0,-4){
\begin{picture}(0,0)(0,0)
\put(0,0){\line(0,1){4}}
\put(4,0){\line(0,1){4}}
\put(4,0){\makebox(8,0)[c]{\ldots}}
\put(12,0){\line(0,1){4}}
\put(16,0){\line(0,1){4}}
\put(0,-1){\makebox(0,0)[t]{\footnotesize \mbox{$1$}\rule[0pt]{0pt}{7pt}}}
\put(4,-1){\makebox(0,0)[t]{\footnotesize \mbox{$2$}\rule[0pt]{0pt}{7pt}}}
\put(12,-1){\makebox(0,0)[t]{\footnotesize
                             \mbox{$\ell\! -\! 1$}\rule[0pt]{0pt}{7pt}}}
\put(16,-1){\makebox(0,0)[t]{\footnotesize \mbox{$\ell$}\rule[0pt]{0pt}{7pt}}}
\end{picture}}
\put(-2,0){\line(1,0){20}}
\put(-2,0){\line(0,1){4}}
\put(-2,4){\line(1,0){20}}
\put(18,0){\line(0,1){4}}
\put(0,4){
\begin{picture}(0,0)(0,0)
\put(0,0){\line(0,1){4}}
\put(4,0){\line(0,1){4}}
\put(4,4){\makebox(8,0)[c]{\ldots}}
\put(12,0){\line(0,1){4}}
\put(16,0){\line(0,1){4}}
\put(0,5){\makebox(0,0)[b]{\footnotesize \mbox{$1'$}}}
\put(4,5){\makebox(0,0)[b]{\footnotesize \mbox{$2'$}}}
\put(12,5){\makebox(0,0)[b]{\footnotesize \mbox{$(\ell\! -\! 1)'$}}}
\put(16,5){\makebox(0,0)[b]{\footnotesize \mbox{$\ell'$}}}
\end{picture}}
\end{picture}}
\put(20,9){\makebox(10,0)[c]{$\longrightarrow$}}
\put(30,0){
\begin{picture}(20,18)(-2,-7)
\put(0,-4){
\begin{picture}(0,0)(0,0)
\put(0,0){\line(0,1){4}}
\put(4,0){\line(0,1){4}}
\put(4,0){\makebox(8,0)[c]{\ldots}}
\put(12,0){\line(0,1){4}}
\put(16,0){\line(0,1){4}}
\put(0,-1){\makebox(0,0)[t]{\footnotesize \mbox{$2$}\rule[0pt]{0pt}{7pt}}}
\put(4,-1){\makebox(0,0)[t]{\footnotesize \mbox{$3$}\rule[0pt]{0pt}{7pt}}}
\put(12,-1){\makebox(0,0)[t]{\footnotesize \mbox{$\ell$}\rule[0pt]{0pt}{7pt}}}
\put(16,-1){\makebox(0,0)[t]{\footnotesize \mbox{$\ell'$}\rule[0pt]{0pt}{7pt}}}
\end{picture}}
\put(-2,0){\line(1,0){20}}
\put(-2,0){\line(0,1){4}}
\put(-2,4){\line(1,0){20}}
\put(18,0){\line(0,1){4}}
\put(0,4){
\begin{picture}(0,0)(0,0)
\put(0,0){\line(0,1){4}}
\put(4,0){\line(0,1){4}}
\put(4,4){\makebox(8,0)[c]{\ldots}}
\put(12,0){\line(0,1){4}}
\put(16,0){\line(0,1){4}}
\put(0,5){\makebox(0,0)[b]{\footnotesize \mbox{$1$}}}
\put(4,5){\makebox(0,0)[b]{\footnotesize \mbox{$1'$}}}
\put(12,5){\makebox(0,0)[b]{\footnotesize \mbox{$(\ell\! -\! 2)'$}}}
\put(16,5){\makebox(0,0)[b]{\footnotesize \mbox{$(\ell\! -\! 1)'$}}}
\end{picture}}
\end{picture}}
\end{picture}
\ee
i.e., one in-string is converted into an out-string and vice versa.
Note that the relative order of the strings is maintained.
The diagrams that one obtains this way can again be interpreted as
representing products of generators in our algebra (no projectors),
although one might have to add additional strings as shown in the example
below.
\be
\rrr{9}{18}
\braidcc{0}{-4}{1}{2}{}{}{}{}
\begin{picture}(0,0)(0,0)
\put(2,0){\oval(4,3.5)[t]}
\put(2,4){\oval(4,3.5)[b]}
\put(0,5){\makebox(0,0)[b]{\footnotesize \mbox{$1'$}}}
\put(4,5){\makebox(0,0)[b]{\footnotesize \mbox{$2'$}}}
\end{picture}
\punit8
\longrightarrow
\punit4
\begin{picture}(0,0)(0,4)
\put(0,0){\line(0,1){3}}
\put(0,5){\line(0,1){3}}
\multiput(0,3)(0.01,0.01){75}{\circle*{0.05}}
\multiput(2,5)(-0.01,-0.01){75}{\circle*{0.05}}
\put(0,5){\line(1,-1){2}}
\put(2,4){\oval(1.75,2)[r]}
\put(4,0){\line(0,1){8}}
\put(0,-1){\makebox(0,0)[t]{\footnotesize \mbox{$2$}\rule[0pt]{0pt}{7pt}}}
\put(4,-1){\makebox(0,0)[t]{\footnotesize \mbox{$2'$}\rule[0pt]{0pt}{7pt}}}
\put(0,9){\makebox(0,0)[b]{\footnotesize \mbox{$1$}}}
\put(4,9){\makebox(0,0)[b]{\footnotesize \mbox{$1'$}}}
\end{picture}
\punit8
\leadsto
\punit4
\vertcr{0}{-6}{2}{}{}
\invbraidcc{0}{-2}{}{}{}{}{}{}
\begin{picture}(0,0)(0,-2)
\put(0,0){\line(0,1){4}}
\put(0,5){\makebox(0,0)[b]{\footnotesize \mbox{$1$}}}
\end{picture}
\punit4
\monoidcc{0}{-6}{3}{3'}{}{}{}{}
\begin{picture}(0,0)(0,-2)
\put(2,0){\oval(4,3.5)[t]}
\put(2,4){\oval(4,3.5)[b]}
\put(0,5){\makebox(0,0)[b]{\footnotesize \mbox{$0$}}}
\put(4,5){\makebox(0,0)[b]{\footnotesize \mbox{$0'$}}}
\end{picture}
\punit4
\vertcr{0}{-2}{}{}{}
\punit4
\begin{picture}(0,0)(0,6)
\put(0,0){\line(0,1){12}}
\put(0,-1){\makebox(0,0)[t]{\footnotesize \mbox{$2'$}\rule[0pt]{0pt}{7pt}}}
\put(0,13){\makebox(0,0)[b]{\footnotesize \mbox{$1'$}}}
\end{picture}
\ee
The labels $(0,0')$ and  $(3,3')$ for the additional lines
are motivated by the fact that the last diagram can also be obtained from
\be
\rrr{7}{15}
\begin{picture}(0,0)(0,4)
\put(0,0){\line(0,1){8}}
\put(0,-1){\makebox(0,0)[t]{\footnotesize \mbox{$0$}\rule[0pt]{0pt}{7pt}}}
\put(0,9){\makebox(0,0)[b]{\footnotesize \mbox{$0'$}}}
\end{picture}
\punit4
\braidcc{0}{-4}{1}{2}{}{}{}{}
\begin{picture}(0,0)(0,0)
\put(2,0){\oval(4,3.5)[t]}
\put(2,4){\oval(4,3.5)[b]}
\put(0,5){\makebox(0,0)[b]{\footnotesize \mbox{$1'$}}}
\put(4,5){\makebox(0,0)[b]{\footnotesize \mbox{$2'$}}}
\end{picture}
\punit8
\begin{picture}(0,0)(0,4)
\put(0,0){\line(0,1){8}}
\put(0,-1){\makebox(0,0)[t]{\footnotesize \mbox{$3$}\rule[0pt]{0pt}{7pt}}}
\put(0,9){\makebox(0,0)[b]{\footnotesize \mbox{$3'$}}}
\end{picture}
\ee
by two generalized crossing steps.

As already mentioned above, we cannot give a natural interpretation
of crossing for the projectors since they only act in one direction.
Nevertheless, we certainly can ``cross'' straight lines that
have a definite colour from the beginning (i.e., one thinks of
the operators acting on coloured strings already which means that
the projectors just act as the identity) to obtain coloured monoids.
For the remaining generators, one has
\be
\vertcr{0}{-1.5}{}{}{}\punit4\vertcr{0}{-1.5}{}{}{}{}
\punit2\longleftrightarrow\punit2
\monoidcc{0}{-1.5}{}{}{}{}{}{}\punit4
\punit4 , \punit4
\braidcc{0}{-1.5}{}{}{}{}{}{}
\punit6\longleftrightarrow\punit2
\invbraidcc{0}{-1.5}{}{}{}{}{}{}\punit4
\ee
i.e., crossing maps the identity to the monoid and
the braid to its inverse.
Thus, by crossing twice, one gets back the diagram one started with.
Note that this is only true if one has only one colour,
otherwise
\be
\vertcl{0}{-1.5}{}{\alpha}{\alpha}\punit4
\vertcr{0}{-1.5}{}{\beta}{\beta}
\punit2\longrightarrow\punit2
\monoidcc{0}{-1.5}{}{}{\beta}{\beta}{\alpha}{\alpha}
\punit6\longrightarrow\punit2
\vertcl{0}{-1.5}{}{\beta}{\beta}\punit4
\vertcr{0}{-1.5}{}{\alpha}{\alpha}
\punit2\longrightarrow\punit2
\monoidcc{0}{-1.5}{}{}{\alpha}{\alpha}{\beta}{\beta}
\punit6\longrightarrow\punit2
\vertcl{0}{-1.5}{}{\alpha}{\alpha}\punit4
\vertcr{0}{-1.5}{}{\beta}{\beta}
\ee
whereas for the coloured braid one still has
\be
\braidcc{0}{-1.5}{}{}{\alpha}{\beta}{\beta}{\alpha}
\punit6\longrightarrow\punit2
\invbraidcc{0}{-1.5}{}{}{\beta}{\alpha}{\alpha}{\beta}
\punit6\longrightarrow\punit2
\braidcc{0}{-1.5}{}{}{\alpha}{\beta}{\beta}{\alpha}
\punit4 \eqstop{.}
\ee
Of course, one always comes back to the operator one started with
after four generalized crossing steps (for $\ell =2$, in general one needs
$2\ell$ steps).

{\em Crossing symmetry} now means that if one has a relation in the algebra
and performs a generalized crossing transformation
on both sides of the equation,
one obtains --- possibly after adding strings
in order to be able to interpret
the resulting diagrams as representing products
of generators of the algebra ---
another valid equation in the algebra. Of course, adding strings
has to be done the same way on both sides of the equation.

To conclude our excursion with an instructive example,
we show how one of the  braid-monoid relations
of eq.~(\ref{BBE}) is crossing-related to the simple fact that
a braid multiplied with its inverse yields the identity.
In the coloured case, it looks as follows
\bea
\rrr{9}{18}
\vertcl{0}{-5}{j}{\beta}{}
\vertcl{0}{-1}{}{}{}
\vertcl{0}{3}{}{}{\beta}
\punit4
\braidcc{0}{-5}{j\! +\! 1}{j\! +\! 2}{\gamma}{\alpha}{}{}
\invbraidcc{0}{3}{}{}{}{}{\gamma}{\alpha}
\vertcl{0}{-1}{}{}{}
\punit4
\vertcl{0}{-1}{}{}{}
\punit4 & = & \punit4
\vertcl{0}{-5}{j}{\beta}{}\vertcl{0}{-1}{}{}{}\vertcl{0}{3}{}{}{\beta}
\punit4
\vertcl{0}{-5}{j\! +\! 1}{\gamma}{}
\vertcl{0}{-1}{}{}{}\vertcl{0}{3}{}{}{\gamma}
\punit4
\vertcr{0}{-5}{j\! +\! 2}{\alpha}{}
\vertcr{0}{-1}{}{}{}\vertcr{0}{3}{}{}{\alpha}
\non \\*
\downarrow\punit8 & & \punit8\downarrow \\*
\rrr{9}{18}
\braidcc{0}{-5}{j}{j\! +\! 1}{\gamma}{\alpha}{}{}
\vertcl{0}{-1}{}{}{}
\monoidcc{0}{3}{}{}{}{}{\beta}{\beta}
\punit4
\braidcc{0}{-1}{}{}{}{}{}{}
\punit4
\vertcr{0}{-5}{j\! +\! 2}{\alpha}{}
\vertcr{0}{3}{}{}{\gamma}
\punit4 & = & \punit4
\vertcl{0}{-5}{j}{\gamma}{}
\vertcl{0}{-1}{}{}{}
\monoidcc{0}{3}{}{}{}{}{\beta}{\beta}
\punit4
\monoidcc{0}{-5}{j\! +\! 1}{j\! +\! 2}{\alpha}{\alpha}{}{}
\vertcl{0}{-1}{}{}{}
\punit4
\vertcr{0}{-1}{}{}{}
\vertcr{0}{3}{}{}{\gamma}
\eea
This also means that one could significantly
reduce the number of defining relations
of the multi-colour algebra by imposing crossing symmetry.
We chose not to do so since it
is not easy to implement crossing algebraically. That also explains
why we used the diagrammatic interpretation of the algebra in the
above discussion.



\section{Representations}
\setcounter{equation}{0}


In the context of exactly solvable models,
we are interested in representations of the multi-colour
braid-monoid algebra where
the central elements $\sq{\alpha}$ and $\om{\alpha}$
($1\leq\alpha\leq m$) are represented by numbers and where
the following equations hold
\bea
\fff{\alpha}{\bb{\alpha}{\alpha}{j}} \;\pp{\alpha}{\alpha}{j}
& = & 0 \label{polyb} \\
\fff{\alpha,\beta}{\bb{\beta}{\alpha}{j}\bb{\alpha}{\beta}{j}} \;
\pp{\alpha}{\beta}{j} & = & 0 \label{polybb} \\
\ggg{\alpha}{\bb{\alpha}{\alpha}{j}}\; \pp{\alpha}{\alpha}{j} & = &
\eee{\alpha}{\alpha}{j} \label{polye}
\eea
where $\fff{\alpha}{z}$, $\fff{\alpha,\beta}{z}$, and $\ggg{\alpha}{z}$
($1\leq\alpha,\beta\leq m$) are polynomials in $z$ and
where we introduced ``two-site projectors'' $\pp{\alpha}{\beta}{j}$
by
\be
\pp{\alpha}{\beta}{j} \;\; = \;\; \PP{\alpha}{j} \PP{\beta}{j+1}
\ee
as a convenient
abbreviation\footnote{Of course, it is possible to use
$\pp{\alpha}{\beta}{j}$ instead $\PP{\alpha}{j}$
to define the algebra from the very beginning. We choose to do otherwise
since in our view, the $\PP{\alpha}{j}$ are the more basic objects.}.
We regard the above relations (\ref{polyb})--(\ref{polye}) as
properties of the representations rather than defining relations
of the algebra since the actual numbers and polynomials are
model-dependent quantities.


The following remarks are in order:
\begin{enumerate}
\item For the one-colour case ($m=1$) the algebra defined above
      reduces to the well-known braid-monoid algebra
      (see e.g.~\cite{WadDegAku})
\item {}From any representation $\varrho$
      of the one-colour algebra
      which acts in an $(n\! +\! 1)$-fold tensor product space
      \be
      \varrho: \;\; \bigotimes_{k=1}^{n+1}\; {\cal V}_{\, k}
               \;\;\;\longrightarrow\;\;\;
               \bigotimes_{k=1}^{n+1}\; {\cal V}_{\, k}
      \label{rep1}
      \ee
      one can obtain
      representations $\varrho^{(m)}$ of the $m$-colour case as follows:
      As representation space at site $j$ choose the $m$-fold direct sum
      $V_{\, j}={\cal V}^{(1)}_{\, j}\oplus{\cal V}^{(2)}_{\, j}
      \oplus\ldots\oplus{\cal V}^{(m)}_{\, j}$
      of the corresponding space of the one-colour representation
      ${\cal V}^{(\alpha)}_{\, j}\cong{\cal V}_{\, j}$
      and represent the operators $\PP{\alpha}{j}$ ($1\leq\alpha\leq m$) by
      the orthogonal projectors onto the $m$
      subspaces ${\cal V}^{(\alpha)}_{\, j}$ at site $j$.
      {}From the compatibility relations (\ref{compb}) and (\ref{compe}),
      it follows immediately that $\bbpm{\alpha}{\beta}{j}$
      and $\eee{\alpha}{\beta}{j}$ can act non-trivially between
      certain subspaces only. Defining
      \bea
      {\bbpm{\alpha}{\beta}{j}}_{\left|\bigotimes_{k}
      {\cal V}^{(\gamma_{k})}_{k}
      \longrightarrow \bigotimes_{k} {\cal V}^{(\delta_{k})}_{k}\right.}
      & = & \left(\prod_{k\neq j,j+1}\delta_{\gamma_{k},\delta_{k}}\right)\;
            \delta_{\alpha,\gamma_{j},\delta_{j+1}}\;
            \delta_{\beta,\gamma_{j+1},\delta_{j}}\; b^{\pm 1}_{\, j}
      \label{rep1b} \\
      {\eee{\alpha}{\beta}{j}}_{\left|\bigotimes_{k}
      {\cal V}^{(\gamma_{k})}_{k}
      \longrightarrow \bigotimes_{k} {\cal V}^{(\delta_{k})}_{k}\right.}
      & = & \left( \prod_{k\neq j,j+1}\delta_{\gamma_{k},\delta_{k}}\right)\;
            \delta_{\alpha,\gamma_{j},\gamma_{j+1}}\;
            \delta_{\beta,\delta_{j},\delta_{j+1}}\; e_{\, j}
      \label{rep1e}
      \eea
      where the products over $k$ run from $1$ to $n+1$ and
      \be
      \delta_{a_{1},a_{2},\ldots,a_{l}}\;\; = \;\;
      \prod_{k=2}^{l}\delta_{a_{1},a_{k}} \eqstop{,}
      \label{mdelta}
      \ee
      one obtains a representation of the $m$-colour braid-monoid
      algebra with $\sq{\alpha}=\sqrt{Q}$, $\om{\alpha}=\omega$,
      $\fff{\alpha}{z}=f(z)$, $\fff{\alpha,\beta}{z} = f(z) f(-z)$, and
      $\ggg{\alpha}{z}=g(z)$ for all $\alpha,\beta=1,\ldots,m$.
      Here, the quantities without colour indices $\alpha$ or
      $\beta$ refer to the one-colour case.
      In particular, this shows that there exist representations
      of the $m$-colour algebra for any $m$.
\item Conversely, starting from a representation of the $m$-colour
      algebra ($m>1$), one can recover part of the one-colour relations for
      the ``full'' or ``uncoloured'' Temperley-Lieb and braid operators
      obtained by summing over all colours. To be more precise,
      the operators $\BBpm{j}$ defined by
      \be
       \BBpm{j} \;\; = \;\;
       \sum_{\alpha,\beta=1}^{m}\; (c^{(\alpha,\beta)})^{\pm 1} \;
       \bbpm{\alpha}{\beta}{j} \label{fullbraid}
      \ee
      with $c^{(\alpha,\beta)}\in\CC\!\setminus\!\{0\}$
      fulfill the braid algebra
      relations (\ref{braidrel}) and the operators $\EE{j}$ defined by
      \be
       \EE{j}  \;\; = \;\;
       \sum_{\alpha,\beta=1}^{m}\;
       \left(\,\frac{c^{(\beta)}}{c^{(\alpha)}}\,\right)\;
       \eee{\alpha}{\beta}{j} \label{fullTL}
      \ee
      ($c^{(\alpha)}\in\CC\!\setminus\!\{0\}$) generate
      the Temperley-Lieb algebra (\ref{TLA}) with
      $\sqrt{Q}=\sum_{\alpha=1}^{m}\sq{\alpha}$.
      However, the relations (\ref{twist}) and (\ref{BBE}) between
      these two types of operators are in general not satisfied by $\BBpm{j}$
      and $\EE{j}$ as they are defined in Eqs.~(\ref{fullbraid}) and
      (\ref{fullTL}) above. In particular,
      the algebra generated by the ``full'' braids and monoids
      in general becomes non-abelian at one site $j$.
      The relations (\ref{BBE}) hold if and only if
      ${(c^{(\alpha,\beta)})}^2=1$ for all $\alpha,\beta=1,\ldots,m$.
      Eq.~(\ref{twist}) is fulfilled with twist $\omega$ if and only if
      $c^{(\alpha,\alpha)}\om{\alpha}=\omega$ for all $\alpha=1,\ldots,m$.
      Hence the full set of relations can only be recovered if all the twists
      $\om{\alpha}$ coincide up to a sign.
      It should be added that whereas one again has polynomial equations in the
      ``full'' braid\footnote{This follows from the fact that there are only
      finitely many independent products of coloured braids which enter
      in the expressions for the powers of the ``full'' braid $\BB{j}$.}
      $\BB{j}$ (\ref{polyb}) one obviously cannot write the
      ``full'' Temperley-Lieb operator $\EE{j}$ as a polynomial
      in the braid  $\BB{j}$ (\ref{polye}) (unless all $\eee{\alpha}{\beta}{j}$
      with $\alpha\neq\beta$ are represented by zero matrices).
\end{enumerate}

In what follows, we construct three classes of representations for the
$m$-colour braid-monoid algebra. These representations are
related to exactly solvable lattice models of statistical mechanics
as we are going to show in sec.~4.


\subsection{Vertex-Type Representations}

Representations of the (one-colour) braid-monoid algebra
which are linked to vertex models (see eg. \cite{WadDegAku})
are of the type
(\ref{rep1}), i.e., they act in a tensor product space.
Therefore, eqs.~(\ref{rep1b}) and (\ref{rep1e}) give $m$-colour
generalizations of this kind of representations.

Here, we construct a different multi-colour generalization
of the representation of the one-colour algebra
related to the 6-vertex model (see eg. \cite{WadDegAku})
which is the simplest (non-trivial) vertex model related
to the affine Lie algebra \VM{A}{1}{1}\ \cite{Jimbo}.
The representation of the $m$-colour braid-monoid algebra
acts in the space
${({\cal V}^{m})}^{\otimes (n+1)}$
with ${\cal V}\cong\CC^{2}$ according to the two possible
states (arrows) of the 6-vertex model. The operators
$\PP{\alpha}{j}$, $\bbpm{\alpha}{\beta}{j}$, and $\eee{\alpha}{\beta}{j}$
act as the identity in all but the space
${({\cal V}^{m})}_{\, j}$ (for $\PP{\alpha}{j}$) respective
${({\cal V}^{m})}_{\, k}$ with $k=j,j+1$
(for $\bbpm{\alpha}{\beta}{j}$ and $\eee{\alpha}{\beta}{j}$),
i.e.,
\bea
\PP{\alpha}{j} & = & \left(\bigotimes_{k=1}^{j-1} 1 \right)\;\otimes\;
\PP{\alpha}{}\;\otimes\;
\left(\bigotimes_{k=j+1}^{n+1} 1 \right) \label{vrep1p} \\
\bbpm{\alpha}{\beta}{j} & = & \left(\bigotimes_{k=1}^{j-1} 1 \right)\;\otimes\;
\bbpm{\alpha}{\beta}{}\;\otimes\;
\left(\bigotimes_{k=j+2}^{n+1} 1 \right) \label{vrep1b}\\
\eee{\alpha}{\beta}{j} & = & \left(\bigotimes_{k=1}^{j-1} 1 \right)\;\otimes\;
\eee{\alpha}{\beta}{}\;\otimes\;
\left(\bigotimes_{k=j+2}^{n+1} 1 \right)
\eqstop{.} \label{vrep1e}
\eea
Here, $\PP{\alpha}{}:\; {\cal V}^{m}\longrightarrow{\cal V}^{m}$ acts
as a projector onto colour $\alpha$
\be
{\PP{\alpha}{}} \;\; = \;\;
\left(\bigoplus_{\beta=1}^{\alpha-1} 0 \right)\;\oplus\; 1
\;\oplus\; \left(\bigoplus_{\beta=\alpha+1}^{m} 0 \right)
\label{vrep2p}
\ee
whereas $\bbpm{\alpha}{\beta}{}$ and $\eee{\alpha}{\beta}{}$
both map
${\cal V}^{m}\otimes{\cal V}^{m}\longrightarrow
 {\cal V}^{m}\otimes{\cal V}^{m}$.
As above (see eqs.~(\ref{rep1b}) and (\ref{rep1e})), the compatibility
relations (\ref{compb}) and (\ref{compe}) imply
that $\bbpm{\alpha}{\beta}{}$ and
$\eee{\alpha}{\beta}{}$ can have non-zero matrix elements
in certain subspaces only. In these subspaces,
they are given by $4\times 4$ matrices with elements
\bea
{\left({\bbpm{\alpha}{\alpha}{}}_{\left|
{\cal V}^{(\alpha)}\otimes{\cal V}^{(\alpha)}
\rightarrow {\cal V}^{(\alpha)}\otimes
{\cal V}^{(\alpha)}\right.}\right)}_{(s,t),(s',t')}
 & = &
{\left( k^{(\alpha)} \right)}^{\pm 1}
\:\left(\;\delta_{s,t'}\;\delta_{s',t} \; - \right. \non \\*
& & \hspace*{1cm} \left.
{(x^{(\alpha)})}^{\pm 1-(s+s')/2}\;\delta_{s,-t}\;
\delta_{s',-t'}\; \right) \label{vrep2b} \\
{\left({\bbpm{\alpha}{\beta}{}}_{\left|
{\cal V}^{(\alpha)}\otimes{\cal V}^{(\beta)}
\rightarrow {\cal V}^{(\beta)}\otimes
{\cal V}^{(\alpha)}\right.}\right)}_{(s,t),(s',t')}
& = & \delta_{s,t'} \; \delta_{s',t}
\hspace*{1cm} \mbox{for $\alpha\neq\beta$}
\label{vrep2bb} \\
{\left({\eee{\alpha}{\beta}{}}_{\left|
{\cal V}^{(\alpha)}\otimes{\cal V}^{(\alpha)}
\rightarrow {\cal V}^{(\beta)}\otimes
{\cal V}^{(\beta)}\right.}\right)}_{(s,t),(s',t')}
 & = &
{\left( x^{(\alpha)}\right)}^{-s}\;{\left( x^{(\beta)}\right)}^{-s'}
\;\delta_{s,-t}\;\delta_{s',-t'}\;
\label{vrep2e}
\eea
where $s,t,s',t'=\pm 1$.
Here, $x^{(\alpha)}=\exp(i\lambda^{(\alpha)})$ and
$k^{(\alpha)}=-i\exp(-i\lambda^{(\alpha)}/2)=1/(i\sqrt{x^{(\alpha)}})$,
and $\lambda^{(\alpha)}$ are arbitrary (real or complex) numbers
which one can choose independently for any colour $\alpha$.

The above representation (which we denote by (6-V,6-V,\ldots,6-V))
is characterized by
\bea
\sq{\alpha}           & = & 2\:\cos(\lambda^{(\alpha)})
                            \;\; = \;\; x^{(\alpha)} + 1/x^{(\alpha)} \non \\
\om{\alpha}           & = & - {\left( k^{(\alpha)} \right)}^{-3} \non \\
\fff{\alpha}{z}       & = & (z - \om{\alpha}) \; (z - k^{(\alpha)})
\label{vertexpoly} \\
\fff{\alpha,\beta}{z} & = & z - 1 \hspace*{1cm} (\alpha\neq\beta) \non \\
\ggg{\alpha}{z}       & = & k^{(\alpha)}\; \left( z - k^{(\alpha)} \right)
\eqstop{.} \non
\eea
Note that even in the case of $m$ identical values of
$\lambda^{(\alpha)}\equiv\lambda$ this representation
is in general different from the representation given in
eqs.~(\ref{rep1b}) and (\ref{rep1e}).
This is obvious
since $\fff{\alpha,\beta}{z}=z-1$ here whereas
$\fff{\alpha,\beta}{z}=(z^{2}-\omega^{2})(z^{2}-k^{2})$ in the
other case.


\subsection{Representations labelled by Graphs}

Consider $m$ (connected) graphs ${\cal G}^{(\alpha)}$, $1\leq\alpha\leq m$,
with $L^{(\alpha)}$ nodes
(enumerated by $a^{(\alpha)}=1,\ldots,L^{(\alpha)}$)
where any pair of nodes is connected by at most one line (bond).
Connectivity is not really a restriction here,
in fact one can always think of the
connected components of one graph as separate graphs.
Note that we do not have to assume that the graphs are simple, i.e.\ that
each line connects two distinct nodes.
In what follows, we denote by
${\cal N}^{(\alpha)}=\{1,\ldots,L^{(\alpha)}\}$ the set of all
nodes of ${\cal G}^{(\alpha)}$.

To each graph ${\cal G}^{(\alpha)}$ we associate an adjacency matrix
${\cal A}^{(\alpha)}$. This $L^{(\alpha)}\times L^{(\alpha)}$ matrix
has elements
\be
{\cal A}^{(\alpha)}_{a^{(\alpha)},b^{(\alpha)}} \;\; = \;\;
{\cal A}^{(\alpha)}_{b^{(\alpha)},a^{(\alpha)}} \;\; = \;\;
\left\{
\ba{l@{\mbox{\hspace*{1cm}}}l}
1 & \mbox{if $a^{(\alpha)}$ and $b^{(\alpha)}$
are adjacent in ${\cal G}^{(\alpha)}$} \\
0 & \mbox{otherwise} \ea \right.
\ee
where $a^{(\alpha)}, b^{(\alpha)}\in\{1,\ldots, L^{(\alpha)}\}$ are
adjacent if the nodes $a^{(\alpha)}$ and $b^{(\alpha)}$ in
${\cal G}^{(\alpha)}$ are connected by a bond.
In other words, the adjacency matrices considered here are characterized
by being symmetric matrices with all entries $0$ or $1$ and
by vanishing elements on the diagonal in case the corresponding
graphs are simple. We denote by
\bea
\Lambda^{(\alpha)}
& = & 2\:\cos(\lambda^{(\alpha)}) \;\; = \;\;
x^{(\alpha)}+1/x^{(\alpha)} \label{para1} \\*
& = & -2\:\cos(4\ol{\lambda}^{(\alpha)}) \;\; = \;\;
      -\left( {(y^{(\alpha)})}^{4} +
{(y^{(\alpha)})}^{-4} \right) \label{para2}
\eea
the Perron-Frobenius eigenvalue of
${\cal A}^{(\alpha)}$ and by $S^{(\alpha)}_{j}$ ($1\leq j\leq L^{(\alpha)}$)
the elements of the corresponding eigenvector, i.e.,
\be
\sum_{i=1}^{L^{(\alpha)}} {\cal A}^{(\alpha)}_{i,j} S^{(\alpha)}_{j}
\;\; =\;\; \Lambda^{(\alpha)} S^{(\alpha)}_{i} \eqstop{.}
\ee
The two different parametrizations in eqs.~(\ref{para1}) and (\ref{para2})
will prove useful below.
Note that whereas for fixed
$\Lambda^{(\alpha)}$ there is
only one value for $\lambda^{(\alpha)}$
(modulo $2\pi$ and up to a sign which is irrelevant here)
there are in general
four different values for $\ol{\lambda}^{(\alpha)}$,
namely $\ol{\lambda}^{(\alpha)}=(\lambda^{(\alpha)}\pm\pi)/4$ and
$\ol{\lambda}^{(\alpha)}=(\lambda^{(\alpha)}\pm 3\pi)/4$.

Consider the cartesian product
${\cal G}={\cal G}^{(1)}\times{\cal G}^{(2)}\times\ldots\times{\cal G}^{(m)}$
of the $m$ graphs. It has $L=\prod_{\alpha=1}^{m}L^{(\alpha)}$ nodes
labelled by
${\cal N}={\cal N}^{(1)}\times{\cal N}^{(2)}\times\ldots\times{\cal N}^{(m)}$
and hence the corresponding adjacency matrix $A$ is an $L\times L$ matrix
given by
\be
A \;\; = \;\; {\cal A}^{(1)}\times{\cal A}^{(2)}
\times\ldots\times{\cal A}^{(m)}
\;\; =\;\; \sum_{\alpha=1}^{m} A^{(\alpha)} \eqstop{,}
\ee
where $A^{(\alpha)}$ is the $L\times L$ matrix with elements
\be
A^{(\alpha)}_{a,b} \;\;= \;\;
\left(\prod_{\beta\neq\alpha} \delta_{a^{(\beta)},b^{(\beta)}}\right)\;
{\cal A}^{(\alpha)}_{a^{(\alpha)},b^{(\alpha)}}
\ee
and we use $m$-tupels $a=(a^{(1)},a^{(2)},\ldots,a^{(m)})\in{\cal N}$
to enumerate the nodes of ${\cal G}$ (which we also call ``states'' or
``heights'' in reminiscence of the role they
play in the solvable models).

The basis $N_{n}$ of our representation space for
the $(n\! +\! 1)$-string $m$-colour braid-monoid
algebra is now given by all allowed
$(n\! +\! 1)$-step paths in the graph ${\cal G}$,
that is by the following set
\be
N_{n} \;\; = \;\;
\left\{ \mbfa =(a_{0},a_{1},\ldots,a_{n},a_{n\! +\! 1}) \; | \;
a_{j}\in {\cal N} \;\mbox{and}\;
\prod_{j=1}^{n+1} A_{a_{j-1},a_{j}} = 1 \right\}
\ee
which forms a subset of the ($n\! +\! 2$)-fold cartesian
product of the set ${\cal N}$.
Then the following equations define a matrix representation of the
$m$-colour algebra on the space spanned by $N_{n}$:
\bea
\left(\PP{\alpha}{j}\;\right)_{\mbfa ,\mbfb} & = &
\left(\prod_{k=0}^{n+1}\delta_{a_{k},b_{k}}\right)\;
A^{(\alpha)}_{a_{j-1},a_{j}}
\label{graphP} \\
\left(\bbpm{\alpha}{\alpha}{j}\;\right)_{\mbfa ,\mbfb} & = &
\left(\prod_{k\neq j}\delta_{a_{k},b_{k}}\right)\;
\left(\;{\left(k^{(\alpha)}\right)}^{\pm 1}\;\delta_{a_{j},b_{j}}\;
A^{(\alpha)}_{a_{j},a_{j+1}}\; A^{(\alpha)}_{a_{j-1},a_{j}}
\;\; +  \right. \non \\*
& & \hspace*{27.5mm} \left.
{\left(k^{(\alpha)}\right)}^{\mp 1}\;
\sqrt{\frac{S_{a_{j}} S_{b_{j}}}{S_{a_{j-1}}^2}}\;
\delta_{a_{j-1},a_{j+1}}\; A^{(\alpha)}_{a_{j},a_{j+1}}\;
A^{(\alpha)}_{a_{j-1},b_{j}}\;
\right) \label{graphbb} \\
\left(\bbpm{\alpha}{\beta}{j}\;\right)_{\mbfa ,\mbfb} & = &
\left(\prod_{k\neq j}\delta_{a_{k},b_{k}}\right)\;
A^{(\alpha)}_{a_{j-1},b_{j}}\; A^{(\beta)}_{b_{j},a_{j+1}}\;
A^{(\alpha)}_{a_{j},a_{j+1}}\; A^{(\beta)}_{a_{j-1},a_{j}}
\mbox{\hspace*{1cm} ($\alpha\neq\beta$)} \label{graphb} \\
\left(\eee{\alpha}{\beta}{j}\;\right)_{\mbfa ,\mbfb} & = &
\left(\prod_{k\neq j}\delta_{a_{k},b_{k}}\right)\;
\delta_{a_{j-1},a_{j+1}}\;
\sqrt{\frac{S_{a_{j}} S_{b_{j}}}{S_{a_{j-1}}^2}}\;
A^{(\alpha)}_{a_{j-1},b_{j}}\; A^{(\beta)}_{a_{j},a_{j+1}}
\label{graphe}
\eea
where $\mbfa =(a_{0},a_{1},\ldots,a_{n\! +\! 1})$,
$\mbfb =(b_{0},b_{1},\ldots,b_{n\! +\! 1})$, and where
\be
k^{(\alpha)} \;\; = \;\; \frac{1}{i\sqrt{x^{(\alpha)}}} \;\; = \;\;
 - {\left(y^{(\alpha)}\right)}^{-2}
\label{scale}\ee
fixes the scale of the braids $\bb{\alpha}{\alpha}{j}$ such
that eqs.~(\ref{BBE}) are satisfied\footnote{Note that
eqs.~(\ref{BBE}) fix the scale of the braids up to a sign.}.

These representations are characterized by
(cf.\ eqs.~(\ref{polyb})--(\ref{polye}))
\bea
\sq{\alpha}           & = & \Lambda^{(\alpha)} \non \\
\om{\alpha}           & = & - {\left( k^{(\alpha)} \right)}^{-3} \non \\
\fff{\alpha}{z}       & = & (z - \om{\alpha}) \; (z - k^{(\alpha)})
\label{RSOSpoly} \\
\fff{\alpha,\beta}{z} & = & z - 1 \hspace*{1cm} (\alpha\neq\beta) \non \\
\ggg{\alpha}{z}       & = & k^{(\alpha)}\; \left( z - k^{(\alpha)} \right)
\eqstop{.}
\non \eea

For the one-colour case, the representation defined through
eqs.~(\ref{graphP})--(\ref{scale}) reduces to the
representation related to the Temperley-Lieb
interaction models \cite{TempLieb,OwczBax,Paul}.
These are critical IRF models \cite{Baxter} if the underlying
graph is a Dynkin diagram of a simply-laced Lie algebra or of
the corresponding affine algebra \cite{Paul}.
They include e.g.\ the
critical \mbox{A--D--E} lattice models of Pasquier \cite{Pas1,Pas2,Pas3}
(which in turn include the ABF or RSOS
(restricted solid-on-solid) models \cite{AndBaxForr})
and the CSOS (cyclic solid-on-solid) models \cite{PeaSea1,PeaSea2,KunYaji}.
For this reason, the present class of representations will frequently be
referred to as ``RSOS representation'' below.

The apparent agreement between the polynomial relations satisfied by
the vertex-type representation (\ref{vertexpoly}) and the present
case (\ref{RSOSpoly}) is not coincidental. If one considers the
representation labelled by
(\VM{A}{L^{(1)}}{1} ,\VM{A}{L^{(2)}}{1},\ldots,\VM{A}{L^{(m)}}{1})
and performs the limits $L^{(\alpha)}\longrightarrow\infty$
with $\lambda^{(\alpha)}=\pi s^{(\alpha)}/L^{(\alpha)}$
($0<s^{(\alpha)}<L^{(\alpha)}$, $s^{(\alpha)}$ and $L^{(\alpha)}$ coprime)
fixed\footnote{Note that in order to do this one has to use other eigenvalues
and eigenvectors of the adjacency matrices
apart from the Perron-Frobenius one. Of course,
one can use any eigenvalue and eigenvector,
the only benefit of using the Perron-Frobenius vector being that
it has real positive entries only.},
one obtains an infinite dimensional representation
of the $m$-colour algebra which nevertheless is related
to the vertex representation (\ref{vrep1p})--(\ref{vrep2e})
of the $(n\! +\! 1)$-string $m$-colour braid monoid algebra
with the corresponding values of $\lambda^{(\alpha)}$.
In the limit, the dependence on the actual heights vanishes ---
all that matters are differences of heights. Since for any colour there are
only two possibilities
(height either increasing or decreasing by one),
one can go over to a description with only two states which
is the corresponding vertex-type representation.
This is nothing but the usual SOS (solid-on-solid) - vertex model
correspondence (Wu-Kadanoff-Wegner transformation,
see \cite{Wu,KadWeg}) which will also be
reflected in our examples of solvable models below.


\subsection{Mixed Representations}

As ``mixed'' representations
(6V,6V,\ldots,6V,${\cal G}^{(1)}$,${\cal G}^{(2)}$,\ldots,${\cal G}^{(m_{R})}$)
we denote representations of the $m$-colour braid-monoid algebra in
which $m_{V}$ colours appear in the vertex-type representation
and the other $m_{R}=m-m_{V}$ in the RSOS representation labelled by the
$m_{R}$ graphs ${\cal G}^{(1)}$, ${\cal G}^{(2)}$, \ldots, ${\cal
G}^{(m_{R})}$.
For simplicity, we assume that among those all graphs are simple.

Due to the mixture of degrees of freedom on vertices and on edges
it is more cumbersome to describe the representation space in the
mixed case.
We use the notation of sec.~3.2 for the colours labelled by the $m_{R}$
simple graphs. For the vertex part, we use the set
\be
{\cal M} \;\; = \;\; \left\{\; s = (s^{(1)},s^{(2)},\ldots,s^{(m_{V})})
\;\left|\;\rule[0mm]{0mm}{4mm}\right.
s^{(\alpha)}\in\{-1,0,1\}\;\;\mbox{and}\;\;
|s|=\sum_{\alpha=1}^{m_{V}} |s^{(\alpha)}|\leq 1\; \right\}
\ee
which gives all possible vertex states on any edge of the lattice.
Here, $0$ stands for no arrow and  $\pm1$ for an arrow that
points upwards (downwards) or to the right (left), respectively.
The second condition guarantees that there is at most one coloured arrow
on any edge.

The representation space $N_{n}$ now consists of all paths
of the form
\bea
N_{n} & = & \left\{\;\rule[0mm]{0mm}{6mm} \mbfa =
(a_{0},s_{1},a_{1},s_{2},a_{2},\ldots,a_{n},s_{n\! +\! 1},a_{n\! +\! 1})
\;\left|\;\rule[0mm]{0mm}{4mm}\right.
a_{j}\in{\cal N},\; s_{j}\in{\cal M}\;\;\mbox{and}\;\; \non \right . \\*
& & \;\;\; \left.\rule[0mm]{0mm}{6mm}
\left(\; A_{a_{j\! -\! 1},a_{j}}\: =\: 1,\;\; |s_{j}|\: =\: 0
\;\right)\;\mbox{or}\;\left(\;\rule[0mm]{0mm}{4mm}
a_{j\! -\! 1}\: =\: a_{j},\;\; |s_{j}|\: =\: 1 \;\right)\;\right\}
\eea
that is, whenever heights on neighbouring vertices
differ by one step on the graph
${\cal G}$, there is no arrow on the bond. On the other hand, if there is a
coloured arrow on any bond, then it has to join two vertices with
equal heights. Other configurations are not allowed.

Now, let us turn to the matrix representations of the generators.
In case they only involve colours of vertex-type, the other
degrees of freedom play no role and the matrices are essentially given by
eqs.~(\ref{vrep1p})--(\ref{vrep2e}). On the other hand, if
only colours of RSOS-type are involved, the vertex degrees of freedom
do not contribute and the relevant parts of the matrices are given by
eqs.~(\ref{graphP})--(\ref{graphe}). Hence, all that remains are the
two-colour braids or monoids where one colour corresponds to a vertex
and the other colour to an RSOS degree of freedom.

These are given explicitly by ($\alpha\neq\beta$)
\bea
{\left(\bbpm{\alpha}{\beta}{j}\right)}_{\mbfa ,\mbfb} & = &
\left(\prod_{k\neq j} \delta_{a_{k},b_{k}}\right) \;
\left(\prod_{k\neq j,j\! +\! 1} \delta_{s_{k},t_{k}}\right)\;
\non \\* & & \;\;\;\;
A^{(\beta)}_{a_{j\! -\! 1},a_{j}}\; A^{(\beta)}_{b_{j},b_{j\! +\! 1}} \;
\delta_{a_{j\! -\! 1},b_{j}}\; \delta_{a_{j},a_{j\! +\! 1}}\;
\delta^{(\alpha)}_{s_{j\! +\! 1},t_{j}}
\label{mixrepb1} \\[4mm]
{\left(\bbpm{\beta}{\alpha}{j}\right)}_{\mbfa ,\mbfb} & = &
\left(\prod_{k\neq j} \delta_{a_{k},b_{k}}\right) \;
\left(\prod_{k\neq j,j\! +\! 1} \delta_{s_{k},t_{k}}\right)\;
\non \\* & & \;\;\;\;
\delta_{a_{j\! -\! 1},a_{j}} \delta_{b_{j},b_{j\! +\! 1}}
A^{(\beta)}_{a_{j\! -\! 1},b_{j}} A^{(\beta)}_{a_{j},a_{j\! +\! 1}}
\delta^{(\alpha)}_{s_{j},t_{j\! +\! 1}}
\label{mixrepb2} \\[4mm]
{\left(\eee{\alpha}{\beta}{j}\right)}_{\mbfa ,\mbfb} & = &
\left(\prod_{k\neq j} \delta_{a_{k},b_{k}}\right) \;
\left(\prod_{k\neq j,j\! +\! 1} \delta_{s_{k},t_{k}}\right)\;
\delta_{a_{j\! -\! 1},b_{j},a_{j\! + \! 1}}\;
\non \\* & & \;\;\;\;
{(x^{(\alpha)})}^{t_{j\! +\! 1}}\;\sqrt{\frac{S_{a_{j}}}{S_{a_{j\! -\! 1}}}}\;
A^{(\beta)}_{a_{j\! -\! 1},a_{j}}\; A^{(\beta)}_{a_{j},a_{j\! +\! 1}}\;
\delta_{-t_{j},t_{j\! +\! 1}}
\label{mixrepe1} \\[4mm]
{\left(\eee{\beta}{\alpha}{j}\right)}_{\mbfa ,\mbfb} & = &
\left(\prod_{k\neq j} \delta_{a_{k},b_{k}}\right) \;
\left(\prod_{k\neq j,j\! +\! 1} \delta_{s_{k},t_{k}}\right)\;
\delta_{a_{j\! -\! 1},a_{j},a_{j\! + \! 1}}\;
\non \\* & & \;\;\;\;
(x^{(\alpha)})^{s_{j\! +\! 1}}\;\sqrt{\frac{S_{b_{j}}}{S_{a_{j\! -\! 1}}}}\;
A^{(\beta)}_{b_{j\! -\! 1},b_{j}} A^{(\beta)}_{b_{j},b_{j\! +\! 1}}
\delta_{-s_{j},s_{j\! +\! 1}}
\label{mixrepe2}
\eea
where $\alpha$ denotes the colour of the vertex-type and
$\beta$ the colour of the RSOS-type degree of freedom.
$\mbfa$ and $\mbfb$ are given by
\bea
\mbfa & = &
(a_{0},s_{1},a_{1},s_{2},a_{2},\ldots,a_{n},s_{n\! +\! 1},a_{n\! +\! 1})
\non \\
\mbfb & = &
(b_{0},t_{1},b_{1},t_{2},b_{2},\ldots,b_{n},t_{n\! +\! 1},b_{n\! +\! 1})
\non
\eea
and $\delta^{(\alpha)}_{s,t}$ is defined as (\ref{mdelta})
\[
\delta^{(\alpha)}_{s,t} \;\; = \;\;
\left(\prod_{\gamma\neq\alpha} \delta_{s^{(\gamma)},t^{(\gamma)},0}\right)\;
\delta_{s^{(\alpha)},t^{(\alpha)}} \eqstop{.}
\]

Of course, the mixed representations are related to the unrestricted
limit of the RSOS representations (for the $m_{V}$ vertex-type
colours) in the same way as the
vertex representations. Hence, also the characteristic polynomial
equations satisfied by the mixed representation are still the
same and are therefore given by eq.~(\ref{RSOSpoly}).



\section{Two-Colour Representations and
         Solutions of the Yang-Baxter Equation}
\setcounter{equation}{0}

Some of the representations of the two-colour braid-monoid algebra
defined in the previous section actually occur in (critical) solvable
RSOS models which are related to (coloured) dense or dilute loop models
(see \cite{WarNieSea,Roche,WarNie}). These models gained considerable
interest after dilute RSOS models labelled by Dynkin diagrams were found
that admit an off-critical extension which breaks the reflection symmetry
of the Dynkin diagram \cite{WarNieSea}. This means that
for these models the elliptic nome
acts as a magnetic field rather than being a temperature-like variable
as it is in the usual case.
As a result, the dilute models related to the Izergin-Korepin
\cite{IzKor,VichResh} (or \VM{A}{2}{2}\ \cite{Jimbo}) vertex model
include, for instance, a model which belongs to the
same universality class as the Ising model in a magnetic field
\cite{WarNieSea}.
We are going to have a closer look at RSOS models which are related to
\VM{A}{2}{1} , \VM{A}{3}{1} , \VM{A}{2}{2} , and \VM{C}{2}{1}\ vertex
models \cite{Jimbo,WarNie}.

The representations of interest for our present purpose are
representations of the two-colour algebra where either
both graphs ${\cal G}^{(1)}$ and ${\cal G}^{(2)}$ coincide
or where one of those
(which without loss of generality we assume to be ${\cal G}^{(2)}$)
is just the Dynkin diagram of the Lie algebra $\mbox{A}_{2}$.
As it turns out, the first case corresponds to RSOS models
related to dense two-colour loop models \cite{WarNie} whereas the
second case corresponds to dilute models \cite{WarNieSea,WarNie}.
To obtain critical models whose Boltzmann weights
are parametrized by trigonometric functions, the graphs
have to be Dynkin diagrams of ``classical'' \mbox{A--D--E}
type or of their affine
counterparts \mbox{A$^{(1)}$--D$^{(1)}$--E$^{(1)}$}.

For the case ${\cal G}^{(2)}=\mbox{A}_{2}$, the above representation
simplifies considerably. The Perron-Frobenius eigenvector
is just given by $S^{(2)}_{1}=S^{(2)}_{2}=1$.
Furthermore, one has $\Lambda^{(2)}=1$ and hence
$x^{(2)}=\exp(\pm \pi i/3)$.
This yields $i\sqrt{x^{(2)}}=\exp(2\pi i/3)$
(resp. $i\sqrt{x^{(2)}}=\exp(\pi i/3)$) and it follows that
\be
\eee{2}{2}{j} \;\; = \;\;
\mp\bb{2}{2}{j} \;\; = \;\;
\mp\bbi{2}{2}{j} \;\; = \;\;
\pp{2}{2}{j}
\label{repA2}
\ee
or, in other words, $\sq{2}=1$, $\om{2}=\mp 1$,
$\fff{2}{z}=z\pm 1$, and $\ggg{2}{z}=\mp z$.
In order to comply with the usual notation
(which is $\lambda^{(\alpha)}=\pi/(L+1)$ for
${\cal G}^{(\alpha)}=\mbox{A}_{L}$)
we choose the first (upper) sign, i.e.\ in particular $\om{2}=-1$.
Actually, since the sign of the braids is not fixed by
eq.~(\ref{BBE}) this is nothing but a convention.

In addition, we discuss several examples of two-colour vertex models
and two-colour mixed vertex-RSOS models. These models are related to
the unrestricted SOS models of the RSOS models mentioned above.
That means in particular that the algebraic description of the
vertex respective vertex-RSOS models is the same as for the
corresponding RSOS models, the only difference being that one deals
with two different representations of the two-colour braid-monoid algebra.

Our algebraic approach clearly shows that the several models listed above,
albeit looking quite different from each other, are in fact closely related
and hence provides a uniform characterization of this type of models.
On the other hand, we are able to show that the Yang-Baxter equations for
these models follow from the algebraic structure alone
(see sec.~4.6). This means that for
any representation of the two-colour algebra with the corresponding properties
one obtains a solution of the Yang-Baxter equations and hence an exactly
solvable model. A more precise formulation as a theorem is presented
at the end of this section.
Finally, the common algebraic structure strongly suggests a
plausible ansatz to find new solvable models. We will come back to this point
in our conclusions.


\subsection{Local Face Operator and the Yang-Baxter Equation}

Let us (very briefly) recall the main properties of exactly
solvable two-dimensional statistical systems which are defined
on a square lattice.
For our purpose, it turns out to be convenient
to think of the square lattice
diagonally as depicted below
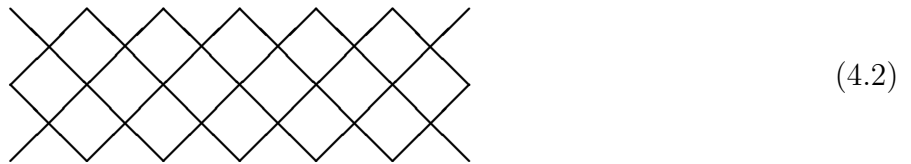
\be
\rrr{7}{14}
\begin{picture}(28,0)(0,0)
\put(0,0){\line(1,1){4}}
\multiput(0,-4)(4,0){5}{\line(1,1){8}}
\put(20,-4){\line(1,1){4}}
\put(4,-4){\line(-1,1){4}}
\multiput(8,-4)(4,0){5}{\line(-1,1){8}}
\put(24,0){\line(-1,1){4}}
\end{picture}
\ee
The models are now defined by specifying the degrees of freedom
(which usually live on the edges or on the vertices of the lattice)
and their interactions.

There are two frequently studied types of models which are
the vertex models and the so-called IRF
(interaction-round-a-face) models.
For vertex models, the degrees of freedom
(``arrows'' or ``spins'')
are located on the edges and the interaction takes
place at the vertices of the lattice.
We denote the Boltzmann weight
of a vertex as follows
\be
\ww{s}{t}{v}{w}{u} \;\; = \;\; \dvertex{s}{t}{v}{w}{u}{0}{0}
\label{VertexWeights}
\ee
where here and in what follows, the letter $u$ is used for
the spectral parameter.
On the other hand, IRF models have their
degrees of freedom (usually called ``heights'') situated
on the vertices. Here, the interaction takes place between the
four corners of an elementary plaquette of the lattice.
The Boltzmann weights for such a plaquette have the form
\be
\WW{a}{b}{c}{d}{u} \;\; =  \;\; \dface{a}{b}{c}{d}{u}{0}{0} \punit4 \eqstop{.}
\label{IRFWeights}
\ee
Of course, one can also consider face models which are
a combination of both. For simplicity, we will not
give explicit expression for these models which can be handled in a
completely analogous way.

We now introduce face transfer operators (also called Yang-Baxter operators)
that act on the configuration along a row of the diagonal lattice.
For IRF models, the face transfer operator $X_{\, j}(u)$
has matrix elements \cite{AkuDegWad88,DegWadAku88b}
\be
{\left( X_{\, j}(u)\right)}_{\mbfap ,\mbfa} \;\; = \;\;
\WW{a_{j}}{a_{j\! +\! 1}}{a'_{j}}{a_{j\! -\! 1}}{u}\;
    \prod_{k\neq j}\delta_{a_{k},a'_{k}} \eqstop{,}
\label{FaceTM}
\ee
where
\bea
\mbfa  & = & (a_{0},a_{1},a_{2},\ldots,a_{n},a_{n\! +\! 1}) \non \\
\mbfap & = & (a'_{0},a'_{1},a'_{2},\ldots,a'_{n},a'_{n\! +\! 1}) \non
\eea
denote allowed height configurations on adjacent rows and
$n$ is the size of of the diagonal lattice. Here, ``allowed'' means
that we consider only those configurations which can be extended to
a configuration on the whole lattice that has a non-zero Boltzmann weight.
For vertex models, the face operators have the following matrix elements
\cite{AkuWad87a,AkuWad87b}
\be
\left(X_{\, j}(u)\right)_{\mbfsp,\mbfs} \;\; = \;\;
\ww{s_{j+1}}{s'_{j+1}}{s'_{j}}{s_{j}}{u}
\;\prod_{k\ne j,j+1}\delta_{s_{k},s'_{k}} \eqstop{.}
\label{VertexTM}
\ee
Here, $\mbfs =(s_{1},s_{2},\ldots,s_{n},s_{n\! +\! 1})$ and
$\mbfsp =(s'_{1},s'_{2},\ldots,s'_{n},s'_{n\! +\! 1})$ denote arrow
configurations along rows of the diagonal lattice.
In fact, the Yang-Baxter operator of a vertex model can
be written as a direct product
\be
X_{\, j}(u) = I\otimes I\otimes\ldots\otimes X(u)\otimes\ldots\otimes I
\ee
where $I$ denotes the identity matrix and
the local operator $X(u)$ covers the $j$ and $j\! +\! 1$
slots.
In both cases, the Yang-Baxter operators $X_{\, j}(u)$ are local operators
in the sense that, except in the vicinity of site $j$, they
act as the identity.

If we assume that the Boltzmann weights
(\ref{VertexWeights}) resp.\ (\ref{IRFWeights})
satisfy the Yang-Baxter equations, the corresponding local face operators
$X_{\, j}(u)$ (eq.~(\ref{FaceTM}) resp.\ eq.~(\ref{VertexTM}))
generate the so-called Yang-Baxter algebra
\cite{Baxter,AkuWad87b,DegWadAku88a}
\renewcommand{\arraystretch}{1.5}
\be
\ba{l}
X_{\, j}(u)\: X_{\, j+1}(u+v)\: X_{\, j}(v) \;\; = \;\;
X_{\, j+1}(v)\: X_{\, j}(u+v)\: X_{\, j+1}(u)  \\
X_{\, j}(u)\: X_{\, k}(v) \;\; = \;\; X_{\, k}(v)\: X_{\, j}(u)
\hspace*{15mm} \mbox{for $|j-k|>1$} \ea
\label{YangBaxterAlgebra}
\ee
\renewcommand{\arraystretch}{1}
The first equation essentially is the Yang-Baxter equation,
the second follows from the local action of the Yang-Baxter operators.
The Yang-Baxter equations guarantee the solvability of the model
since it follows that the row transfer matrices of the model form a
commuting one-parameter family which yields an infinite number
of conserved quantities.
Usually, one requires that the Boltzmann weights have
additional properties besides fulfilling the Yang-Baxter equations.
Two of them, the standard initial condition
(value of Boltzmann weights at $u=0$)
and the inversion relation (unitarity condition), can be formulated
directly in terms of the face transfer operators. They result in the
relations
\renewcommand{\arraystretch}{1.5}
\be
\ba{l@{\hspace*{10mm}}l}
X_{\, j}(0) \;\; = \;\; \id &
\mbox{\footnotesize (Standard Initial Condition)} \\
X_{\, j}(u)\: X_{j}(-u) \;\; = \;\; \varrho(u)\:\varrho(-u)\:\id &
\mbox{\footnotesize (Inversion Relation)} \ea
\label{InversionRelation}
\ee
\renewcommand{\arraystretch}{1}
where $\varrho(u)$ is a model-dependent
function of the spectral parameter.
Note that the inversion relation can actually be derived from the
Yang-Baxter equation and the standard initial condition.

Another important property of many exactly solvable models is
the so-called crossing symmetry. Since it cannot be
simply described in terms
of the local face operators, we go back to the Boltzmann weights.
For IRF models, crossing symmetry means
\be
\WW{a}{b}{c}{d}{u} \;\; = \;\;
{\left(\frac{\psi(a)\,\psi(c)}{\psi(b)\,\psi(d)}\right)}^{1/2}\;
\WW{b}{c}{d}{a}{\lambda-u}
\label{CrossIRF}
\ee
and for vertex models, one has
\be
\ww{s}{t}{v}{w}{u} \;\; = \;\;
{\left(\frac{r(s)\, r(w)}{r(t)\, r(v)}\right)}^{1/2}\;
\ww{t}{\ol{v}}{w}{\ol{s}}{\lambda-u} \eqstop{.}
\label{CrossVertex}
\ee
Here, $\lambda$ is the crossing parameter and $\psi(a)$ resp.\ $r(s)$ are
crossing multipliers \cite{AkuDegWad88,AkuWad87b,DegWadAku88a}
which are complex numbers and do not depend on the spectral parameter.
For the vertex models, $\ol{s}=-s$ means the charge conjugated state of $s$
and the crossing multipliers satisfy $r(\ol{s})=1/r(s)$.
The presence of crossing multipliers means that the Boltzmann weights
are not invariant under rotations in general.
However, the crossing multipliers of
adjacent vertices resp.\ plaquettes cancel each
other and on a periodic lattice
(which means periodic in both directions)
they do not enter into the partition
function at all. This is an example of the more general concept
of gauge equivalence. Two models are gauge-equivalent if their
Boltzmann weights are related by a gauge transformation.
In the present context, these are
transformations of the Boltzmann weights which do not change
the partition function of the model on a lattice which is periodic in
both directions. A general local gauge transformation
therefore has the form
\be
\WW{a}{b}{c}{d}{u} \;\;\;\longmapsto\;\;\;
\frac{F(d,c)\, G(d,a)}{F(a,b)\, G(c,b)}\;\:\WW{a}{b}{c}{d}{u}
\label{GaugeIRF}
\ee
for IRF models and
\be
\ww{s}{t}{v}{w}{u} \;\;\;\longmapsto\;\;\;
\frac{F(w,v)\, G(w,s)}{F(s,t)\, G(v,t)}\;\:\WW{s}{t}{v}{w}{u}
\label{GaugeVertex}
\ee
for vertex models. In both cases, $F$ and $G$ can be arbitrary functions
which also are allowed to depend on the spectral parameter $u$.
Of course, quantities which are not determined by the partition function
on the periodic lattice alone may depend on the particular gauge.
For instance, the Yang-Baxter algebra of a solvable model can depend
on the gauge. For an example, see the discussion of the \VM{A}{2}{2}\
vertex model below.


\subsection{RSOS Models labelled by the pair
            (\mbox{\protect\boldmath $\cal G$},A$_{2}$)}


The two examples of solvable RSOS models we want to consider here
are related to the \VM{A}{2}{1}\
resp.\ \VM{A}{2}{2}\ vertex models \cite{Jimbo}.
As these models are defined in refs.~\cite{WarNieSea} and \cite{WarNie},
they are labelled by one {\em simple} graph ${\cal G}$
(which is of \mbox{A--D--E} type for critical models) and in addition the
height variable is allowed to stay at the same value for adjacent sites,
i.e.\ the actual adjacency diagram $\tilde{\cal G}$ is obtained from
${\cal G}$ by affixing a circle at each node of ${\cal G}$.
For ${\cal G}=\mbox{A}_{L}$ for instance,
$\tilde{\cal G}$ thus looks as follows:
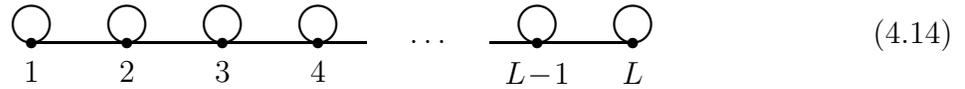
\be
\rrr{4}{9}
\begin{picture}(17.5,0)(0,0)
\multiput(0,0)(5,0){4}{\circle*{0.5}}
\multiput(0,1)(5,0){4}{\circle{2}}
\put(0,0){\line(1,0){17.5}}
\put(0,-1){\makebox(0,0)[t]{\mbox{$1$}}}
\put(5,-1){\makebox(0,0)[t]{\mbox{$2$}}}
\put(10,-1){\makebox(0,0)[t]{\mbox{$3$}}}
\put(15,-1){\makebox(0,0)[t]{\mbox{$4$}}}
\end{picture}
\punit2
\ldots
\punit2
\begin{picture}(7.5,0)(-2.5,0)
\put(-2.5,0){\line(1,0){7.5}}
\multiput(0,0)(5,0){2}{\circle*{0.5}}
\multiput(0,1)(5,0){2}{\circle{2}}
\put(0,-1){\makebox(0,0)[t]{\mbox{$L\! -\!1$}}}
\put(5,-1){\makebox(0,0)[t]{\mbox{$L$}}}
\end{picture}
\ee

In order to interpret this model as an RSOS model labelled by a pair
$({\cal G},\mbox{A}_{2})$ of graphs, we double the number of heights.
That is instead $a\in{\cal G}$ we have two heights
$(a,b)\in ({\cal G},\mbox{A}_{2})$ with $b\in\{1,2\}$ and hence we have
two copies of the graph ${\cal G}$. The circles are now replaced
by lines connecting the heights $(a,1)$ and $(a,2)$ in the two different
copies. Thus what we get is the graph ${\cal G}\times\mbox{A}_{2}$
which for our above example
${\cal G}=\mbox{A}_{L}$ looks as follows:
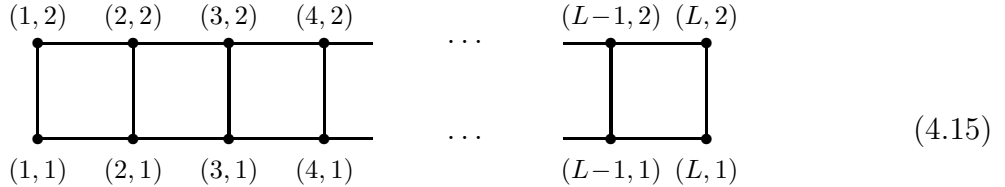
\be
\rrr{4}{13}
\begin{picture}(35,0)(0,0)
\multiput(0,0)(5,0){4}{\circle*{0.5}}
\multiput(0,5)(5,0){4}{\circle*{0.5}}
\multiput(0,0)(5,0){4}{\line(0,1){5}}
\put(0,0){\line(1,0){17.5}}
\put(0,5){\line(1,0){17.5}}
\put(27.5,0){\line(1,0){7.5}}
\put(27.5,5){\line(1,0){7.5}}
\multiput(30,0)(5,0){2}{\circle*{0.5}}
\multiput(30,5)(5,0){2}{\circle*{0.5}}
\multiput(30,0)(5,0){2}{\line(0,1){5}}
\put(17.5,0){\makebox[10\unitlength]{\mbox{\ldots}}}
\put(17.5,5){\makebox[10\unitlength]{\mbox{\ldots}}}
\put(0,-1){\makebox(0,0)[t]{\footnotesize \mbox{$(1,1)$}}}
\put(5,-1){\makebox(0,0)[t]{\footnotesize\mbox{$(2,1)$}}}
\put(10,-1){\makebox(0,0)[t]{\footnotesize\mbox{$(3,1)$}}}
\put(15,-1){\makebox(0,0)[t]{\footnotesize\mbox{$(4,1)$}}}
\put(30,-1){\makebox(0,0)[t]{\footnotesize\mbox{$(L\! -\!1,1)$}}}
\put(35,-1){\makebox(0,0)[t]{\footnotesize\mbox{$(L,1)$}}}
\put(0,6){\makebox(0,0)[b]{\footnotesize\mbox{$(1,2)$}}}
\put(5,6){\makebox(0,0)[b]{\footnotesize\mbox{$(2,2)$}}}
\put(10,6){\makebox(0,0)[b]{\footnotesize\mbox{$(3,2)$}}}
\put(15,6){\makebox(0,0)[b]{\footnotesize\mbox{$(4,2)$}}}
\put(30,6){\makebox(0,0)[b]{\footnotesize\mbox{$(L\! -\!1,2)$}}}
\put(35,6){\makebox(0,0)[b]{\footnotesize\mbox{$(L,2)$}}}
\end{picture}
\ee
The graph ${\cal G}\times\mbox{A}_{2}$ possesses a natural
$\mbox{\ZZ}_{2}$ symmetry ($(a,b)\longmapsto (a,\ol{b})$)
with $\ol{b}=3-b$ and by taking the quotient
\be
\left( {\cal G}\times\mbox{A}_{2}\right) /\mbox{\ZZ}_{2}
\;\;\cong\;\; \tilde{\cal G}
\ee
one recovers the graph $\tilde{\cal G}$. This, of course,
is a rather trivial example of the orbifold duality of
Fendley and Ginsparg \cite{FenGin}.

There is an exact one-to-two correspondence between
``$\tilde{\cal G}$-allowed'' faces
and ``${\cal G}\times\mbox{A}_{2}$-allowed'' faces
\be
\rrr{5}{10}
\dface{a}{b}{c}{d}{}{0}{0}\punit8 \longleftrightarrow \punit6
\dface{(a,e)}{(b,\ol{e})}{(c,e)}{(d,\ol{e})}{}{0}{0} \punit4
\ee
with $e\in\{1,2\}$ and $\ol{e}=3-e$.
It is therefore obvious that for any
``$\tilde{\cal G}$-allowed'' configuration on a square lattice there are
exactly two ``${\cal G}\times\mbox{A}_{2}$''-allowed configurations. Hence
choosing the face weights to be
\be
\mbox{\ww{(a,1)}{(b,2)}{(c,1)}{(d,2)}{u}} \;\; = \;\;
\mbox{\ww{(a,2)}{(b,1)}{(c,2)}{(d,1)}{u}} \;\; = \;\;
\mbox{\ww{a}{b}{c}{d}{u}} \eqstop{,}
\ee
one obtains a solvable model with adjacency
graph ${\cal G}\times\mbox{A}_{2}$
whose partition function is just twice the
partition function of the corresponding dilute model with
adjacency graph $\tilde{\cal G}$.


As our first example, we consider the dilute model
which is related to the \VM{A}{2}{1}\
vertex model \cite{WarNie}.
The local face operator (\ref{FaceTM}) for this model
can be written as follows
\bea
X_{j}(u)\;\; & = & \;\;\slu \;\left(\pp{1}{1}{j} + \pp{2}{2}{j}\right) \;
 + \; \left( \pp{1}{2}{j} + \pp{2}{1}{j}\right)  \non \\*
& & + \;\su\; \left(\eee{1}{1}{j} + \bb{1}{2}{j} + \bb{2}{1}{j} \right)
\label{A21RSOS}
\eea
where the braid and monoid operators are understood to be
the matrices of the representation
labelled by $({\cal G},\mbox{A}_{2})$ and
$\lambda=\lambda^{(1)}$.
Obviously, the model is not crossing symmetric
since $\pp{1}{2}{j}$ and $\pp{2}{1}{j}$ enter in $X_{j}(u)$
with a coefficient one whereas
the corresponding ``crossed'' objects
$\eee{1}{2}{j}$ and $\eee{2}{1}{j}$ do not show up in the
expression for $X_{j}(u)$ at all.


The second example is the dilute model related to the \VM{A}{2}{2}\
vertex model (Izergin-Korepin model) \cite{WarNieSea,WarNie}.
Its Yang-Baxter operator (\ref{FaceTM}) has the form
\bea
X_{j}(u) \;\; & = & \;\;\slulu{2}{3}{2}{3}\; \pp{1}{1}{j}
\;\; + \;\;\left( 1 + \sulu{3}{2}{3} \right)\; \pp{2}{2}{j} \non \\
& & + \;\sslu{3}{3}\;\left( \pp{1}{2}{j} + \pp{2}{1}{j} \right)
\label{A22RSOS} \\
& & - \;\sulu{ }{2}{3}\; \eee{1}{1}{j} \;
    + \;\ssu{3}\; \left(\eee{1}{2}{j} + \eee{2}{1}{j}\right) \non \\
& & + \;\sulu{3}{2}{3}\;\left(\bb{1}{2}{j} + \bb{2}{1}{j}\right) \non
\eea
with $\ol{\lambda}=\ol{\lambda}^{(1)}$.

As one recognizes from this expression
for the local face operator $X_{j}(u)$,
the face weights are manifestly crossing symmetric.
The crossing parameter of the model is $3\ol{\lambda}$.

We choose this model as our
main example since it is particularly interesting.
Not only does the
${\cal G}=\mbox{A}_{L}$ model with $L$ odd allow an off-critical
extension which breaks the reflection symmetry of the
Dynkin diagram $\mbox{A}_{L}$ \cite{WarNieSea}, but
the Yang-Baxter algebra (\ref{YangBaxterAlgebra}) of the critical model
is also non-commuting at one site:
\be
\left[ X_{j}(u), X_{j}(v) \right] \;\; \neq \;\; 0 \eqstop{.}
\ee
Of course, this implies that the ``full'' braid $\BBpm{j}$ obtained from
\bea
\BBpm{j} & = & - y^{\mp 2} \lim_{u\rightarrow \mp i\infty}
                     \frac{X_{j}(u)}{\varrho(u)} \non \\
& = & - y^{\mp 2} \pp{1}{1}{j} - y^{\pm 2} \eee{1}{1}{j} +
             \left(\pp{2}{2}{j} + \bb{1}{2}{j} + \bb{2}{1}{j}\right)
\label{fullb} \\
& = & \bbpm{1}{1}{j} + \bbpm{1}{2}{j} + \bbpm{2}{1}{j} - \bbpm{2}{2}{j} \non
\eea
with $y=\exp (i\ol{\lambda})$,
$\bb{2}{2}{j}=-\pp{2}{2}{j}$ (cf.\ eq.~(\ref{repA2}))
and the ``full'' monoid $\EE{j}$ which is given by
\be
\EE{j} \;\; = \;\; X_{j}(3\ol{\lambda}) \;\; = \;\;
\eee{1}{1}{j} + \eee{1}{2}{j} + \eee{2}{1}{j} + \pp{2}{2}{j} \;\; = \;\;
\eee{1}{1}{j} + \eee{1}{2}{j} + \eee{2}{1}{j} + \eee{2}{2}{j} \label{fulle}
\ee
do not commute. Here,
\be
\varrho(u) \;\; = \;\; \slulu{2}{3}{2}{3}
\ee
is the function that enters in the inversion
relation (\ref{InversionRelation}).
It is therefore obvious that this model cannot be
described in terms of the usual (one-colour) braid-monoid algebra.
Furthermore, in the general case we were not able to
find a gauge transformation (\ref{GaugeIRF})
of the face weights which results in an equivalent model
(i.e., a model with the same partition function on a torus)
with a commuting algebra at one
site\footnote{We found such a gauge for the model labelled by
${\cal G}=\mbox{A}^{(1)}_{L}$, compare the remarks on the gauge of
the \VM{A}{2}{2}\ vertex model below.}.

It is not possible to express the local face operator
$X_{\, j}(u)$ in terms of the ``full'' braids and monoids
alone. One has to add at least one additional operator,
which for instance one can choose to be
\be
F_{\, j}\;\; =\;\;\pp{1}{1}{j}+\eee{1}{1}{j}+
\pp{2}{2}{j}+\eee{2}{2}{j}
\;\; = \;\; \pp{1}{1}{j} + \eee{1}{1}{j} + \; 2\pp{2}{2}{j}
\eqstop{.} \label{fullf}
\ee
This yields
\bea
X_{\, j}(u) \;\;
& = & \;\; \sslu{3}{3}\;\:\id \;\;\; +\;\;\;\ssu{3}\;\:\EE{j} \non \\
&   & + \; \frac{\sin(\frac{u}{2})\:\sin(\frac{3\ol{\lambda} - u}{2})}
                {\sin(2\ol{\lambda})\:\sin(3\ol{\lambda})}\;\:
           \left\{\;\cos{(\frac{3\ol{\lambda}}{2})}\;
           \left(\:\BB{j}\: +\:\BBi{j}\:\right) \right. \non \\*
&   &  \hspace*{39mm} \left. \; +\; \rule{0mm}{6mm}
       \left(\: e^{-i\, (\frac{3\ol{\lambda}}{2}-u)}\:\BB{j}\: +\:
                e^{i\, (\frac{3\ol{\lambda}}{2}-u)}\:\BBi{j}\:\right)\;\right\}
       \label{fXA22}  \\
&   & + \; \left(\; 1\; -\;\sslu{3}{3}\; -\;\ssu{3}\; \right)\; F_{\, j}
\eqstop{.}\non
\eea

The relations for the ``full'' braids and monoids
and the additional operator $F_{\, j}$ can be
computed from the relations for the coloured operators.
As it turns out,
the ``full'' braid $\BB{j}$ (\ref{fullb}) satisfies a quartic equation
\be
\left(\BB{j} - \id\right)\;\left(\BB{j} + \id\right)\;
\left(\BB{j} + y^{-2} \id\right)\;
\left(\BB{j} - y^{6} \id\right) \;\; = \;\; 0
\ee
whereas the ``full'' monoid $\EE{j}$ (\ref{fulle}) fulfills the quadratic
\be
{\left(\EE{j}\right)}^{2}
\;\; = \;\; \left(1 - y^{4} - y^{-4}\right)\;\EE{j}
\;\; = \;\; \left(1 - 2\:\cos(4\ol{\lambda})\right)\;\EE{j}
\ee
and the additional operator $F_{\, j}$ (\ref{fullf})
satisfies another quartic
\be
F_{\, j} \left(F_{\, j} -\id\right)\; \left(F_{\, j} - 2 \id\right)\;
\left(F_{\, j} -(1-y^{4}-y^{-4})\id\right) \;\; = \;\; 0 \eqstop{.}
\ee
The commutator of the braid and the monoid,
\be
C_{\, j}\;\; =\;\;\frac{1}{y^{\pm 6}-1}\;\left[\BBpm{j},\EE{j}\right]
\eqstop{,}
\ee
fulfills a simple cubic relation
\be
C_{\, j}\; \left( C_{\, j} - \id \right) \;\left( C_{\, j} + \id \right)
\;\; =\;\; 0 \eqstop{.}
\ee
Furthermore, one finds the following commutation relations
\be
\left[ F_{\, j},\BBpm{j} \right] \;\; = \;\; 0
\hspace*{4mm} , \hspace*{4mm}
\left[ F_{\, j},\EE{j} \right] \;\; = \;\;
-\left( 1 + y^4 + y^{-4} \right) \; C_{\, j}
\eqstop{,}
\ee
to present just a few of the many relations that one
can derive from the two-colour algebra.

So far, we only considered some
relations of operators at the same site.
It turns out that the ``full'' braids and monoids in fact fulfill
all defining relations of the one-colour braid-monoid algebra except
the twist relations.
In particular, the one-colour braid-monoid
relations (\ref{BBE}) are satisfied.
This actually follows from the third remark we made
in the beginning of this section
(see eqs.~(\ref{fullbraid}) and (\ref{fullTL})).
However, we want to emphasize again that neither
$\EE{j}$ nor $F_{\, j}$ is a polynomial in the braid $\BB{j}$.
To describe the algebra completely
in terms of $\BB{j}$, $\EE{j}$, and $F_{\, j}$,
one needs of course additional relations with operators
acting on tho neighbouring sites which
involve the operator $F_{\, j}$ and which can be derived from
the corresponding relations of the two-colour algebra.

As we demonstrated,
one can alternatively describe the local Yang-Baxter algebra
of the dilute model
in terms of the (spectral-parameter independent) operators
$\BB{j}$, $\EE{j}$, and $F_{\, j}$.
However, it should be clear from this exercise
that the formulation using the two-colour
braid-monoid algebra has several advantages.
It is simpler and certainly  more natural in the sense that the two-colour
algebra is a straight-forward generalization of the Temperley-Lieb algebra
and that it still allows a graphical interpretation and therefore is related
to properties of coloured knots and links.


\subsection{RSOS Models labelled by the pair
(\mbox{\protect\boldmath $\cal G$},\mbox{\protect\boldmath $\cal G$})}

Again, we present two examples one of which consists of a series
of crossing symmetric models.
The models considered here are the dense two-colour loop models of
ref.~\cite{WarNie}. We are going to have a closer look at the
RSOS models related to \VM{C}{2}{1}\ since these models
are crossing symmetric.


First, let us have a short glance at an IRF model related
to the \VM{A}{3}{1}\ vertex model \cite{WarNie}. The
Yang-Baxter operator (\ref{FaceTM}) for the model
reads as follows
\bea
X_{j}(u) \;\; & = & \;\;\slu\;
\left(\pp{1}{1}{j} + \pp{2}{2}{j}\right) \;
 + \;\left( \pp{1}{2}{j} + \pp{2}{1}{j}\right) \non \\*
& & + \;\su\; \left(\eee{1}{1}{j} + \eee{2}{2}{j} +
      \bb{1}{2}{j} + \bb{2}{1}{j} \right) \eqstop{,}\label{A31RSOS}
\eea
where $\lambda=\lambda^{(1)}=\lambda^{(2)}$ is determined
by eq.~(\ref{para1}).
As the models defined by eq.~(\ref{A21RSOS}) above,
this model does not possess a crossing symmetry.
Again, the two-colour monoids
$\eee{\alpha}{\beta}{j}$ ($\alpha\neq\beta$) do not enter
in the expression for the local face operator at all.


We are more interested in our second example
which is the above mentioned IRF model related to the
\VM{C}{2}{1}\  vertex model \cite{WarNie}.
The local face operator (\ref{FaceTM}) of this
model has the form
\bea
X_{j}(u)\;\; & = & \;\;\slulu{2}{6}{2}{6}\;
\left(\pp{1}{1}{j}+\pp{2}{2}{j}\right)
\;\; + \;\;\sslu{6}{6}\; \left( \pp{1}{2}{j} + \pp{2}{1}{j}\right) \non \\
& & - \;\sulu{4}{2}{6}\;\left(\eee{1}{1}{j} + \eee{2}{2}{j}\right) \;
    + \;\ssu{6}\; \left(\eee{1}{2}{j} + \eee{2}{1}{j}\right) \non \\
& & + \;\sulu{6}{2}{6}\;\left(\bb{1}{2}{j} + \bb{2}{1}{j}\right)
\label{C21RSOS}
\eea
with $\ol{\lambda}=\ol{\lambda}^{(1)}=\ol{\lambda}^{(2)}$
(cf.\ eq.~(\ref{para2})).
Here, the crossing symmetry is manifestly built-in again,
with crossing parameter $6\ol{\lambda}$.

In contrast to the previous crossing-symmetric example
(\ref{A22RSOS}), the local face operator is still commuting
at one site, i.e.,
\be
\left[ X_{j}(u), X_{j}(v) \right] \;\; = \;\; 0 \eqstop{.}
\ee
This stems from the fact that the model is labelled
by two copies of the same graph and hence the twists
$\om{1}$ and $\om{2}$ coincide which guarantees the commutativity
of the ``full'' braid and monoid. Still, one cannot express the
local face operator $X_{\, j}(u)$ entirely in terms of
the ``full'' braids and monoids. One again has to introduce
(at least) one additional operator, which
in the present case can be chosen as
\be
F_{\, j} \;\; = \;\; \bb{1}{2}{j} + \bb{2}{1}{j} \eqstop{,}
\ee
for instance. For the local face operator, one obtains
\bea
X_{\, j}(u) \;\;
& = & \;\; \sslu{6}{6}\;\:\id \;\;\; +\;\;\;\ssu{6}\;\:\EE{j} \non \\
&   & + \; \frac{\sin(\frac{u}{2})\:\sin(\frac{6\ol{\lambda} - u}{2})}
                {\sin(2\ol{\lambda})\:\sin(6\ol{\lambda})}\;\:
           \left\{\;\frac{\cos{(\ol{\lambda})}}{\cos{(2\ol{\lambda})}}\;
           \left(\:\BB{j}\: +\:\BBi{j}\:\right) \right. \non \\*
&   &  \hspace*{39mm} \left. \; +\; \rule{0mm}{6mm}
       \left(\: e^{-i\, (\frac{6\ol{\lambda}}{2}-u)}\:\BB{j}\: +\:
       e^{i\, (\frac{6\ol{\lambda}}{2}-u)}\:\BBi{j}\:\right)\;\right\} \\
&   & + \; \frac{1}{2\,\cos{(2\ol{\lambda})}}\;
       \left(\; 1\; -\;\sslu{6}{6}\; -\;\ssu{6}\; \right)\; F_{\, j} \non
\eea
which shows striking similarity to eq.~(\ref{fXA22}).
Here, the ``full'' braids and monoids are given by
\bea
\BBpm{j} & = & -y^{\mp2}\;
\lim_{u\rightarrow\mp i\infty} \frac{X_{\, j}(u)}{\varrho (u)}
\;\; = \;\;
\sum_{\alpha,\beta}\; \bbpm{\alpha}{\beta}{j} \non \\
\EE{j}   & = &
X_{\, j}(6\ol{\lambda}) \;\; = \;\;
\sum_{\alpha,\beta}\; \eee{\alpha}{\beta}{j} \non
\eea
where
\be
\varrho (u) \;\; = \;\; \slulu{2}{6}{2}{6}
\ee
is the function that enters in the inversion relation
(\ref{InversionRelation}). The following relations hold
\renewcommand{\arraystretch}{1.5}
\be
\ba{l}
\displaystyle
\left(\BB{j} - \id\right)\;\left(\BB{j} + \id\right)\;
\left(\BB{j} + y^{-2} \id\right)\;
\left(\BB{j} - y^{6} \id\right) \;\; = \;\; 0 \\
\displaystyle
{\left(\EE{j}\right)}^{2}
\;\; = \;\; -2\;\left(y^{4} + y^{-4}\right)\;\EE{j}
\;\; = \;\; -4\:\cos{(4\ol{\lambda})}\;\EE{j} \\
\displaystyle
\BB{j}\;\EE{j} \;\; = \;\; \EE{j}\;\BB{j} \;\; = \;\; y^{6}\; \EE{j} \\
\displaystyle
F_{\, j}\;\left(F_{\, j} + \id\right)\; \left(F_{\, j} - \id\right)
\;\; = \;\; 0 \\
\displaystyle
\left[\:\BBpm{j}\: ,\:F_{\, j}\:\right] \;\; = \;\;
\left[\:\EE{j}\: ,\:F_{\, j}\:\right] \;\; = \;\; 0 \eqstop{.}
\ea
\label{C21rel}
\ee
\renewcommand{\arraystretch}{1}
Obviously, this very much resembles a representation of the one-colour
braid-monoid algebra where the braid satisfies a
fourth order polynomial equation. However, the monoid $\EE{j}$
as well as the additional operator $F_{\, j}$ cannot be written
as a polynomial in the braid $\BB{j}$.


\subsection{Vertex Models of type (6-V,6-V)}

We are going to mention two models which fall into
this category: the \VM{A}{3}{1}\ and \VM{C}{2}{1}\ vertex models
\cite{Jimbo}. These are of course equivalent to the unrestricted case
of the two series of RSOS models defined above.

This implies that the Yang-Baxter operators of
the \VM{A}{3}{1}\ and \VM{C}{2}{1}\ vertex models
are given by the very same expressions as
those of the related face models, i.e.\ by
eq.~(\ref{A31RSOS}) for the \VM{A}{3}{1}\ vertex model and
by eq.~(\ref{C21RSOS}) for the \VM{C}{2}{1}\ case.
The only difference is that the representation of the two-colour
braid-monoid algebra is now of vertex-type (6V,6V)
with $\lambda^{(1)}=\lambda^{(2)}=\lambda$
and $\ol{\lambda}=(\lambda\pm\pi)/4$ or
$\ol{\lambda}=(\lambda\pm 3\pi)/4$.
It is given explicitly by
eqs.~(\ref{vrep1p})--(\ref{vrep2e}) above.

Of course, the results on the ``full'' braids and monoids
for the RSOS models related to \VM{C}{2}{1}\
(see eqs.~(\ref{C21RSOS})--(\ref{C21rel})) also carry over
to the vertex model.
This may surprise many readers since the Yang-Baxter
algebra of the \VM{C}{2}{1}\ vertex model usually is known to be
a Birman-Wenzl-Murakami (BWM) algebra \cite{BirWen,Mura},
hence a one-colour
braid-monoid algebra where the braids satisfy a third order
polynomial equation. However, the vertex model we consider
here differs from the \VM{C}{2}{1}\ vertex model of Jimbo
\cite{Jimbo} by a spectral-parameter dependent gauge transformation
(\ref{GaugeVertex})
which affects the braid limit and therefore influences the
algebraic relations. This is well-known, see e.g.\
eq.~(6.12) of ref.~\cite{WadDegAku}
where such a gauge transformation is used in order to obtain
an ``interesting'' braid group representation from the six-vertex model.
The same happens in the case of the \VM{A}{2}{2} vertex model which
is going to be discussed subsequently.


\subsection{Mixed Vertex-RSOS Models}

Here, we again have two examples which are linked to
the RSOS models considered above, namely the
\VM{A}{2}{1}\ and \VM{A}{2}{2}\ vertex models \cite{Jimbo}
which are related to representations of type
(6-V,$\mbox{A}_{2}$).
As above,
these models are equivalent to the unrestricted case
of the two corresponding series of RSOS models
defined in eqs.~(\ref{A21RSOS}) and (\ref{A22RSOS}).
Hence the expressions for the Yang-Baxter
operator of the \VM{A}{2}{1}\ and \VM{A}{2}{2}\ vertex
models again coincide with those of the related face models
which are given in eq.~(\ref{A21RSOS}) for
the \VM{A}{2}{1}\ models and in eq.~(\ref{A22RSOS}) for the
\VM{A}{2}{2}\ models, respectively. The corresponding
representation of the two-colour braid-monoid algebra
is now the mixed representation which we called (6-V,$\mbox{A}_{2}$)
(see eqs.~(\ref{mixrepb1})--(\ref{mixrepe2}) above).

For the \VM{A}{2}{2}\ vertex model,
also known under the name Izergin-Korepin model
\cite{IzKor,VichResh}, the same comment applies as for the
\VM{C}{2}{1}\ vertex model before. Again, the vertex model
obtained as the unrestricted case of the dilute A--D--E models
differs from the \VM{A}{2}{2}\ vertex model of
refs.~\cite{IzKor,Jimbo,VichResh} by a spectral-parameter
dependent gauge (\ref{GaugeVertex}).
Of course, we posed ourselves the question
if one can get a one-site commuting Yang-Baxter algebra for the
dilute A--D--E models by an appropriate
spectral-parameter dependent gauge transformation
(\ref{GaugeIRF}). It turns out
that this is at least possible for the dilute \VM{A}{L}{1} models
where the gauge transformation of the vertex model carries over
directly. For the other cases, notably for the most interesting
case of the dilute $\mbox{A}_{L}$ models, we did not succeed to
find such a transformation
(although we also tried to use the additional gauge freedom that one
gains by the doubling of the states we did in order to
interpret the dilute models as two-colour models)
and it is our opinion that at least a
local gauge of the form (\ref{GaugeIRF}) does not exist.

Recent results of S.~O.~Warnaar and B.~Nienhuis \cite{WarNie}
suggest that there are in fact several infinite series of
mixed vertex-RSOS models.
These include models with
an arbitrary number of colours,
but all of them (except one or two) occur in the vertex-type
representation. As an explicit example,
they consider the \VM{A}{3}{2}\ loop-vertex model which in our language
would correspond to a mixed vertex-RSOS model of type
(6-V,$\cal G$) which however is not crossing symmetric.

In \cite{WarNie}, these models are obtained by a partial mapping
of well-known vertex models \cite{Jimbo} to loop models.
The loop degrees of freedom in turn can be rewritten as an
RSOS model (in the sense that their partition functions on an infinite
lattice coincide, for more details see \cite{WarNie}).
The vertex models investigated in \cite{WarNie} in this context
are the \VM{A}{n}{1} , \VM{A}{n}{2} , and \VM{C}{n}{1}\
vertex models \cite{Jimbo}.


\subsection{Yang-Baxterization}

So far, we presented several examples
of critical exactly solvable models whose
local Yang-Baxter operators can be written in terms of
two-colour braid-monoid operators in certain matrix representations.
Now, we address the question to which extent the algebraic
structure is responsible for the solvability of the model,
i.e.\ in particular for the inversion relation
and the Yang-Baxter equation. Let us formulate the result
as a theorem:

\begin{theorem}
Let $\PP{\alpha}{j}$, $\bbpm{\alpha}{\beta}{j}$, and
$\eee{\alpha}{\beta}{j}$ ($\alpha = 1,2$)
denote a representation of
the two-colour braid-monoid algebra
and $\pp{\alpha}{\beta}{j}=\PP{\alpha}{j}\PP{\beta}{j}$.
Then the following statements hold:
\begin{enumerate}
\item If $\sq{1}=2\cos(\lambda)$,
         $\fff{1,2}{z}=\fff{2,1}{z}=z-1$,
      and if $X_{\, j}(u)$ is given by eq.~(\ref{A21RSOS}),
      then $X_{\, j}(u)$ generates a Yang-Baxter algebra
      (\ref{YangBaxterAlgebra}) and
      the inversion relation (\ref{InversionRelation})
      holds with $\varrho(u)=\slu$.
\item If $\sq{1}=\sq{2}=2\cos(\lambda)$,
         $\fff{1,2}{z}=\fff{2,1}{z}=z-1$,
      and if $X_{\, j}(u)$ is given by eq.~(\ref{A31RSOS}),
      then $X_{\, j}(u)$ generates a Yang-Baxter algebra
      (\ref{YangBaxterAlgebra}) and
      the inversion relation (\ref{InversionRelation})
      holds with $\varrho(u)=\slu$.
\item If $\sq{1}=-2\cos(4\ol{\lambda})$,
         $\sq{2}=1$,
         $\fff{1,2}{z}=\fff{2,1}{z}=z-1$,
         $\fff{2}{z}=z+1$,
         $\ggg{2,2}{z}=-z$,
      and if $X_{\, j}(u)$ is given by eq.~(\ref{A22RSOS}),
      then $X_{\, j}(u)$ generates a Yang-Baxter algebra
      (\ref{YangBaxterAlgebra}) and
      the inversion relation (\ref{InversionRelation})
      holds  with $\varrho(u)=\slulu{2}{3}{2}{3}$.
\item If $\sq{1}=\sq{2}=-2\cos(4\ol{\lambda})$,
         $\fff{1,2}{z}=\fff{2,1}{z}=z-1$,
      and if $X_{\, j}(u)$ is given by eq.~(\ref{C21RSOS}),
      then $X_{\, j}(u)$ generates a Yang-Baxter algebra
      (\ref{YangBaxterAlgebra}) and
      the inversion relation (\ref{InversionRelation})
      holds with $\varrho(u)=\slulu{2}{6}{2}{6}$.
\end{enumerate}
\end{theorem}
The proofs consist of lengthy, but straight-forward
direct calculations. Note that the values of $\om{\alpha}$
and the functions $\fff{\alpha}{z}$
do not always enter since we eliminated all braids
$\bbpm{\alpha}{\alpha}{j}$
in the expressions for the Yang-Baxter operator $X_{j}(u)$.
Nevertheless, they are fixed indirectly, since
performing a braid limit on these expressions determines
the braids in terms of the other operators (up to
normalization).



\section{Conclusions}
\setcounter{equation}{0}


We defined a multi-colour braid-monoid algebra
as a straight-forward
generalization of the one-colour case.
The diagrammatic interpretation also generalizes from the
one-colour case. This means that there is a direct relation to
the theory of coloured knots and links.
The notion of crossing symmetry was discussed using the pictorial
representation of the algebra.
Different general classes of representations
which are connected to solvable lattice models
were given explicitly. The two-colour algebra turned out to
describe the Yang-Baxter algebra of recently constructed
solvable models which are related to dilute one-colour loop models
\cite{WarNieSea,Roche,WarNie}
and to dense two-colour loop models \cite{WarNie}.
For several examples (which were taken from ref.~\cite{WarNie}),
we expressed the Yang-Baxter operator of the models
in terms of generators of the two-colour braid-monoid
algebra and our main result is that we could
actually prove that the solvability of the model
follows from the algebraic structure alone. These include
the so-called dilute A--D--E models \cite{WarNieSea,Roche}
which are of particular interest since the dilute A$_{L}$
models with $L$ odd allow an integrable off-critical
extension that breaks the symmetry of the
Dynkin diagram A$_{L}$. Hence, the physical meaning of
the elliptic nome is that of a magnetic field.
This has for instance the consequence that,
although the two-dimensional Ising model in a magnetic field has not
been solved, the dilute A$_{3}$ model provides us with
an exactly solvable model which belongs to the same universality class.

The dilute A--D--E models possess another interesting property.
Their Yang-Baxter algebra, which is the algebra
generated by the local face operators, is non-commuting at one site.
For these models, a description using only ``full'' braids
and monoids (which are obtained essentially by summing up the
coloured objects) turned out to be complicated.
It involved at least one new object and we doubt that
there is a nice diagrammatic interpretation for the relations
satisfied by these operators. Our description of the dilute
models as two-colour models is certainly more natural.
For the rest of our examples, similar observations apply.


As one notices, there is an apparent similarity in the expressions for the
Yang-Baxter operators of the
RSOS models labelled by $({\cal G},\mbox{A}_{2})$
(eqs.~(\ref{A21RSOS})--(\ref{A22RSOS})) and of
those labelled by $({\cal G},{\cal G})$
(eqs.~(\ref{A31RSOS})--(\ref{C21RSOS})).
This suggests that it might be possible to Yang-Baxterize \cite{Jones}
any representation of the two-colour braid-monoid algebra
with the same properties as the representation labelled by any pair
$({\cal G}^{(1)},{\cal G}^{(2)})$ of graphs.
\begin{conjecture}
Let $\pp{\alpha}{\beta}{j}$, $\bbpm{\alpha}{\beta}{j}$,
and $\eee{\alpha}{\beta}{j}$ ($\alpha,\beta=1,2$) be a
representation of the two-colour braid monoid algebra
with
\bea
\sq{\alpha}           & = & - {\left(k^{(\alpha)}\right)}^{2} \; - \;
                            {\left(k^{(\alpha)}\right)}^{-2} \non \\
\om{\alpha}           & = & - {\left( k^{(\alpha)} \right)}^{-3} \non \\
\fff{\alpha}{z}       & = & (z - \om{\alpha}) \; (z - k^{(\alpha)})  \\
\fff{\alpha,\beta}{z} & = & z - 1 \hspace*{1cm} (\alpha\neq\beta) \non \\
\ggg{\alpha}{z}       & = & k^{(\alpha)}\; \left( z - k^{(\alpha)} \right)
\eqstop{,} \non
\eea
where $k^{(1)}$ and $k^{(2)}$ are arbitrary complex numbers (of modulus one).
Then this representation can be Yang-Baxterized to a (critical)
solvable model whose Yang-Baxter operator is of the form
\bea
X_{j}(u) & = & \;\;\; f_{1}(u)  \pp{1}{1}{j}  + f_{2}(u)  \pp{2}{2}{j} +
                      f_{3}(u)  \pp{1}{2}{j}  + f_{4}(u)  \pp{2}{1}{j} \non \\
         &   & + \;   f_{5}(u)  \eee{1}{1}{j} + f_{6}(u)  \eee{2}{2}{j} +
                      f_{7}(u)  \eee{1}{2}{j} + f_{8}(u)  \eee{2}{1}{j}  \\
         &   & + \;   f_{9}(u)  \bb{1}{1}{j}  + f_{10}(u) \bb{2}{2}{j} +
                      f_{11}(u) \bb{1}{2}{j}  + f_{12}(u) \bb{2}{1}{j} \non
\eea
where the coefficient functions $f_{i}(u)$ are products of
trigonometric functions of the spectral parameter
$u$ involving (in general) both $\lambda^{(1)}$ and $\lambda^{(2)}$.
\end{conjecture}
Even more, one might conjecture that this extends to the $m$-colour case
although we are not aware of any known restricted model which is related to
the $m$-colour algebra with $m>2$.
Work to check these conjectures is currently in progress.


In case these conjectures turn out to be correct, this would provide
us with a method to obtain new solvable critical models.
In some sense this procedure resembles the fusion
(both have a diagrammatic interpretation acting on
composite strings, cf.\ e.g.\ ref.~\cite{WadDegAku}) of solvable models
but, as the above examples show clearly, it in fact leads to totally
different models.

Furthermore, all examples we presented in this paper
in fact correspond to two-colour
generalizations of the Temperley-Lieb algebra only.
In particular, this means that the coloured braids fulfill
quadratic equations.
One would certainly expect that one can find representations
of the multi-colour algebra and solvable models related to them
which for instance generalize
the BWM algebra (where the braids fulfill a cubic equation).
Hence our results open up a variety of interesting directions
for further investigations.


The authors gratefully acknowlegde financial support from
the Deutsche Forschungsgemeinschaft (UG) and from the
Australian Research Council.
We would like to thank B.~Nienhuis,
K.~A.~Seaton and S.~O.~Warnaar for helpful comments and
for sending us their results prior to publication.
One of us (PAP) would like to thank B.~Nienhuis and FOM for hospitality
and support during his visit to the University of Amsterdam.



\begin{thebibliography}{99}

{\small

\bibitem{Jimbo89}
M.~Jimbo,
\newblock {\em Yang-Baxter Equation in Integrable Systems},
\newblock  World Scientific, Singapore, 1989

\bibitem{YangGe89}
C.~N.~Yang and M.~L.~Ge,
\newblock {\em Braid Group, Knot Theory and Statistical Mechanics},
\newblock  World Scientific, Singapore, 1989

\bibitem{TempLieb}
H.~N.~V.~Temperley and E.~H.~Lieb,
\newblock Proc. Roy. Soc. (London) A322 (1971) 251

\bibitem{Martin91}
P.~P.~Martin,
\newblock {\em Potts Models and Related Problems in Statistical Mechanics},
\newblock World Scientific, Singapore, 1991

\bibitem{Kauffman}
L.~Kauffman,
\newblock Topology 26 (1987) 395

\bibitem{BirWen}
J.~Birman and H.~Wenzl,
\newblock preprint (1987); Trans. Am. Math. Soc. 313 (1989) 249

\bibitem{Mura}
J.~Murakami,
\newblock Osaka J. Math. 24 (1987) 745

\bibitem{WadDegAku}
M.~Wadati, T.~Deguchi and Y.~Akutsu,
\newblock  Phys. Rep. 180 (1989) 247

\bibitem{WarNie}
S.~O.~Warnaar and B.~Nienhuis,
{\em Solvable lattice models labelled by Dynkin diagrams},
\newblock preprint

\bibitem{WarNieSea}
S.~O.~Warnaar, B.~Nienhuis and K.~A.~Seaton,
\newblock Phys. Rev. Lett. 69 (1992) 710

\bibitem{Roche}
Ph.~Roche,
\newblock {\em On the construction of integrable dilute \mbox{A--D--E} models},
\newblock preprint

\bibitem{Jones}
V.~F.~R.~Jones,
\newblock Int. J. Mod. Phys. B4 (1990) 701

\bibitem{AkuDeg91}
Y.~Akutsu and T.~Deguchi,
\newblock Phys. Rev. Lett. 67 (1991) 777

\bibitem{DegAku91a}
T.~Deguchi and Y.~Akutsu,
\newblock J. Phys. Soc. Japan 60 (1991) 2559

\bibitem{Deg}
T.~Deguchi,
\newblock J. Phys. Soc. Japan 60 (1991) 3978

\bibitem{DegAku91b}
T.~Deguchi and Y.~Akutsu,
\newblock J. Phys. Soc. Japan 60 (1991) 4051

\bibitem{AkuDegWad92}
Y.~Akutsu, T.~Deguchi and T.~Wadati,
\newblock J. Knot Theory and Ram. 1 (1992) 161

\bibitem{CheGeLiuXue}
Y.~Cheng, M.-L.~Ge, G.~C.~Liu and K.~Xue,
\newblock J. Knot Theory and Ram. 1 (1992) 31

\bibitem{Jimbo}
M.~Jimbo,
\newblock Commun. Math. Phys. 102 (1986) 537

\bibitem{OwczBax}
A.~L.~Owczarek and R.~J.~Baxter,
\newblock J. Stat. Phys. 49 (1987) 1093

\bibitem{Paul}
P.~A.~Pearce,
\newblock Int. J. Mod. Phys. B4 (1990) 715

\bibitem{Baxter}
R.~J.~Baxter,
\newblock {\em Exactly Solved Models in Statistical Mechanics},
\newblock Academic Press, London, 1982

\bibitem{Pas1}
V.~Pasquier,
\newblock Nucl. Phys. B285 (1987) 162

\bibitem{Pas2}
V.~Pasquier,
\newblock J. Phys. A20 (1987) 5707

\bibitem{Pas3}
V.~Pasquier,
\newblock J. Phys. A20 (1987) L1229

\bibitem{AndBaxForr}
G.~E.~Andrews, R.~J.~Baxter and P.~J.~Forrester,
\newblock J. Stat. Phys. 35 (1984) 193

\bibitem{PeaSea1}
P.~A.~Pearce and K.~A.~Seaton,
\newblock Phys. Rev. Lett. 60 (1988) 1347

\bibitem{PeaSea2}
P.~A.~Pearce and K.~A.~Seaton,
\newblock Ann. Phys. (N.Y.) 193 (1989) 326

\bibitem{KunYaji}
A.~Kuniba and T.~Yajima,
\newblock J. Stat. Phys. 52 (1988) 829

\bibitem{Wu}
F.~Y.~Wu,
\newblock Phys. Rev. B4 (1971) 2312

\bibitem{KadWeg}
L.~P.~Kadanoff and J.~Wegner,
\newblock Phys. Rev. B4 (1971) 3983

\bibitem{IzKor}
A.~G.~Izergin and V.~E.~Korepin,
\newblock Commun. Math. Phys. 79 (1981) 303

\bibitem{VichResh}
V.~I.~Vichirko and N.~Yu.~Reshetikhin,
\newblock Teor. i Mat. Fiz. 56 (1983) 260

\bibitem{AkuDegWad88}
Y.~Akutsu, T.~Deguchi and M.~Wadati,
\newblock J. Phys. Soc. Japan 57 (1988) 1173

\bibitem{DegWadAku88b}
T.~Deguchi, M.~Wadati and Y.~Akutsu,
\newblock J. Phys. Soc. Japan 57 (1988) 2921

\bibitem{AkuWad87a}
Y.~Akutsu and M.~Wadati,
\newblock J. Phys. Soc. Japan 56 (1987) 839

\bibitem{AkuWad87b}
Y.~Akutsu and M.~Wadati,
\newblock J. Phys. Soc. Japan 56 (1987) 3039

\bibitem{DegWadAku88a}
T.~Deguchi, M.~Wadati and Y.~Akutsu,
\newblock J. Phys. Soc. Japan 57 (1988) 1905

\bibitem{FenGin}
P.~Fendley and P.~Ginsparg,
\newblock Nucl. Phys. B324 (1989) 549
}

\end{thebibliography}
\end{document}